\documentclass[9pt,a4paper,onecolumn,twoside]{tau}
\usepackage[english]{babel}

\usepackage{umlaute,eurosym,comment}
\usepackage{amsmath,amssymb,amsthm,bm}
\usepackage{float,epsfig,pstricks,pst-node,pst-bezier,auto-pst-pdf,xcolor}
\usepackage{wrapfig}
\usepackage{multirow,dcolumn}
\usepackage{tikz,pgfplots}
\usepackage{soul}
\usepackage{hyperref}
\psset{unit=1mm,framesep=10pt,fillstyle=none}  % pstricks basic settings
\degrees[360]                                  % segmentation of the unit circle
\psset{linewidth=0.9pt}                        % standard line width
\providecommand{\mel}{\psset{linewidth=0.5pt}} % medium line
\providecommand{\stl}{\psset{linewidth=0.9pt}} % standard line
\providecommand{\bol}{\psset{linewidth=1.4pt}} % bold line

\newlength\figureheight	
\newlength\figurewidth	

\providecommand{\ef}{\;.}
\providecommand{\ec}{\;,}

\providecommand{\fa}{\forall\;}         % for all
\providecommand{\tp}{^{\scriptsize{\mathrm{T}}}} % transpose
\providecommand{\to}{\rightarrow}                        % converges to
\providecommand{\opt}{^{\star}}                          % optimal element
\newcommand{\BN}{\mathbb{N}}        % natural numbers
\newcommand{\BZ}{\mathbb{Z}}        % integers
\newcommand{\BR}{\mathbb{R}}        % real numbers
\providecommand{\autompc}{\texttt{AutoMPC}}
\providecommand{\acknlmpc}{\texttt{ackermann\_nlmpc}}

\definecolor{ULOcean}{RGB}{0,75,90}
\definecolor{ULOrange}{RGB}{236,116,4}
\definecolor{LiteRed}{RGB}{240,170,170}
\definecolor{MedRed}{RGB}{204,24,24}
\definecolor{DarkRed}{RGB}{127,15,15}
\definecolor{LiteYellow}{RGB}{255,245,128}
\definecolor{MedYellow}{RGB}{255,237,32}
\definecolor{DarkYellow}{RGB}{255,235,0}
\definecolor{LiteBlue}{RGB}{170,184,255}
\definecolor{MedBlue}{RGB}{43,78,255}
\definecolor{DarkBlue}{RGB}{0,42,120}
\definecolor{LiteGreen}{RGB}{207,224,207}
\definecolor{MedGreen}{RGB}{40,120,40}
\definecolor{DarkGreen}{RGB}{0,80,0}
\definecolor{LiteViolet}{RGB}{207,195,220}
\definecolor{MedViolet}{RGB}{80,29,127}
\definecolor{DarkViolet}{RGB}{53,0,107}
\definecolor{LiteOrange}{RGB}{255,220,187}
\definecolor{MedOrange}{RGB}{255,150,50}
\definecolor{DarkOrange}{RGB}{255,124,0}
\definecolor{LiteBrown}{RGB}{225,192,180}
\definecolor{MedBrown}{RGB}{154,40,0}
\definecolor{DarkBrown}{RGB}{116,30,0}
\definecolor{LiteGray}{RGB}{230,230,230}
\definecolor{MedGray}{RGB}{160,160,160}
\definecolor{DarkGray}{RGB}{100,100,100}

\definecolor{taucolor}{rgb}{0.0,0.2,0.4}       % Blue
\definecolor{taucolornote}{rgb}{0.0,0.2,0.4}       % Blue, Background / Remark
\definecolor{taucolorexam}{rgb}{0.12, 0.3, 0.17}   % Green, Example
\definecolor{taucolorinst}{rgb}{0.3, 0.13, 0.26}   % Red, Instruction

\journalname{Technical Documentation}
\title{AutoMPC: A Code Generator for MPC-based Automated Driving}

\author[*]{Georg Schildbach}
\author[*]{Jasper Pflughaupt}
\affil[*]{Autonomous Systems Laboratory, University of Luebeck}
\professor{Version 1.0 -- August 19, 2025\vspace*{0.3cm}\\ Contact: gschildbach@gmail.com}

\institution{University of Luebeck}
\footinfo{Version 1.0}
\theday{August 19, 2025}
\leadauthor{Schildbach and Pflughaupt}
\course{Documentation \texttt{AutoMPC}}

\begin{abstract}    
    Model Predictive Control (MPC) is a powerful technique to control nonlinear, multi-input  multi-output systems subject to input and state constraints. It is now a standard tool for trajectory tracking control of automated vehicles. As such it has been used in many research and development projects.
	However, MPC faces several challenges to be integrated into industrial production vehicles. The most important ones are its high computational demands and the complexity of implementation. 
    The software packages \autompc\; aims to address both of these challenges. It builds on a robustified version of an active set algorithm for Nonlinear MPC. The algorithm is embedded into a framework for vehicle trajectory tracking, which makes it easy to used, yet highly customizable. Automatic code generation transforms the selections into a standalone, computationally efficient C-code file with static memory allocation. As such it can be readily deployed on a wide range of embedded platforms, e.g., based on Matlab/Simulink\textregistered\; or Robot Operating System (ROS).
    Compared to a previous version of the code, the vehicle model and the numerical integration method can be manually specified, besides basic algorithm parameters. All of this information and all specifications are directly baked into the generated C-code. The algorithm is suitable driving scenarios at low or high speeds, even drifting, and supports direction changes.
    Multiple simulation scenarios show the versatility and effectiveness of the \autompc\; code, with the guarantee of a feasible solution, a high degree of robustness, and computational efficiency.
\end{abstract}


\begin{document}
		
    \maketitle 
    \thispagestyle{firststyle} 
    \tauabstract 
    % \tableofcontents
    % \linenumbers 
    
%----------------------------------------------------------

\section{Introduction}

\taustart{M}odel Predictive Control (MPC) has become a popular method for automated driving. Major advantages include its natural ability to control nonlinear multi-input multi-output systems, include predictions, and satisfy constraints on the states and inputs. The key drawback remains its computational complexity and, perhaps even more importantly, its relative difficulty of implementation.\\
The software package \autompc\; aims to fill this gap, by providing a monolithic, fully operational MPC block for automated driving---instead of a generic optimization solver that must first be adapted. \autompc\; generates robust and efficient C code, which can be used directly in Matlab/Simulink\textregistered or Robot Operating System (ROS). Despite its simplicity, the MPC is highly configurable, including the choice of an internal vehicle model, the integration method, and solver parameters.\\
This technical report provides the necessary instructions for the installation and usage of the \autompc\; package, as well as some technical details and background information. 

\subsection{Automated Driving Architecture} 

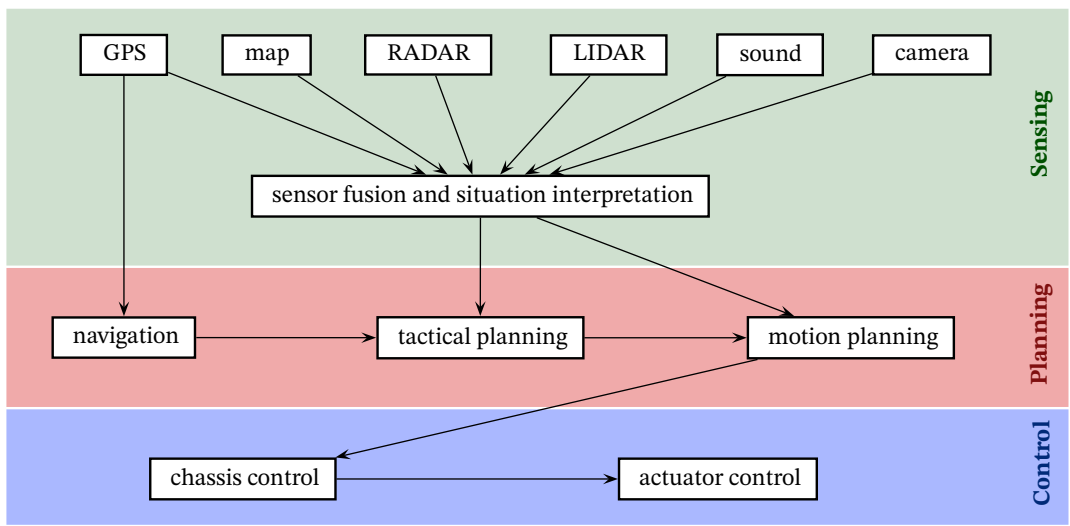
\begin{figure}[H]
  \centering
		\psset{unit=1.25mm}
     \begin{pspicture}(-15,-5)(97,50)
         % Background
        \psframe[linestyle=none,fillstyle=solid,fillcolor=LiteGreen](-15,50)(97,22.5)
        \psframe[linestyle=none,fillstyle=solid,fillcolor=LiteRed](-15,22.5)(97,7.5)
        \psframe[linestyle=none,fillstyle=solid,fillcolor=LiteBlue](-15,7.5)(97,-5)
        \rput{90}(94,36.25){\psframebox[linestyle=none]{\normalsize\textcolor{DarkGreen}{\textbf{Sensing}}\normalsize\hspace*{-0.1cm}}}
        \rput{90}(94,15){\psframebox[linestyle=none]{\normalsize\textcolor{DarkRed}{\textbf{Planning}}\normalsize\hspace*{-0.1cm}}}
        \rput{90}(94,1.25){\psframebox[linestyle=none]{\normalsize\textcolor{DarkBlue}{\textbf{Control}}\normalsize\hspace*{-0.1cm}}}
		%\scriptsize
		\stl
		\psset{arrowsize=4pt}
		\rput[bc](-2.5,45){\rnode{S1}{\psframebox[framesep=3pt,fillstyle=solid,fillcolor=white]{\phantom{p}GPS
		\phantom{P}\hspace*{-0.1cm}}}}
		\rput[bc](12.5,45){\rnode{S2}{\psframebox[framesep=3pt,fillstyle=solid,fillcolor=white]{\phantom{p}map
		\phantom{P}\hspace*{-0.1cm}}}}
		\rput[bc](29.5,45){\rnode{S3}{\psframebox[framesep=3pt,fillstyle=solid,fillcolor=white]{\phantom{p}RADAR
		\phantom{P}\hspace*{-0.1cm}}}}
		\rput[bc](48.5,45){\rnode{S4}{\psframebox[framesep=3pt,fillstyle=solid,fillcolor=white]{\phantom{p}LIDAR
		\phantom{P}\hspace*{-0.1cm}}}}
		\rput[bc](65.5,45){\rnode{S5}{\psframebox[framesep=3pt,fillstyle=solid,fillcolor=white]{\phantom{p}sound
		\phantom{P}\hspace*{-0.1cm}}}}
		\rput[bc](82.5,45){\rnode{S6}{\psframebox[framesep=3pt,fillstyle=solid,fillcolor=white]{\phantom{p}camera
		\phantom{P}\hspace*{-0.1cm}}}}
		\rput[bc](35,30){\rnode{LO}{\psframebox[framesep=3pt,fillstyle=solid,fillcolor=white]{\phantom{p}\hspace*{-0.1cm}
		sensor fusion and situation interpretation\phantom{P}\hspace*{-0.1cm}}}}
		\rput[bc](-2.5,15){\rnode{NV}{\psframebox[framesep=3pt,fillstyle=solid,fillcolor=white]{\phantom{p}\hspace*{-0.1cm}
		navigation\phantom{P}\hspace*{-0.1cm}}}}
		\rput[bc](35,15){\rnode{TP}{\psframebox[framesep=3pt,fillstyle=solid,fillcolor=white]{\phantom{p}\hspace*{-0.1cm}
		tactical planning\phantom{P}\hspace*{-0.1cm}}}}
		\rput[bc](74,15){\rnode{PP}{\psframebox[framesep=3pt,fillstyle=solid,fillcolor=white]{\phantom{p}\hspace*{-0.1cm}
		motion planning\phantom{P}\hspace*{-0.1cm}}}}
		\rput[bc](10,0){\rnode{VC}{\psframebox[framesep=3pt,fillstyle=solid,fillcolor=white]{\phantom{p}\hspace*{-0.1cm}
		chassis control\phantom{P}\hspace*{-0.1cm}}}}
		\rput[bc](60,0){\rnode{AC}{\psframebox[framesep=3pt,fillstyle=solid,fillcolor=white]{\phantom{p}\hspace*{-0.1cm}
		actuator control\phantom{P}\hspace*{-0.1cm}}}}
		\mel
		\ncline[nodesep=0pt]{->}{S1}{LO}
		\ncline[nodesep=0pt]{->}{S2}{LO}
		\ncline[nodesep=0pt]{->}{S3}{LO}
		\ncline[nodesep=0pt]{->}{S4}{LO}
		\ncline[nodesep=0pt]{->}{S5}{LO}
		\ncline[nodesep=0pt]{->}{S6}{LO}
		\ncline[nodesep=0pt]{->}{S1}{NV}
		\ncline[nodesep=0pt]{->}{NV}{TP}
		\ncline[nodesep=0pt]{->}{LO}{TP}
		\ncline[nodesep=0pt]{->}{LO}{PP}
		\ncline[nodesep=0pt]{->}{TP}{PP}
		\ncline[nodesep=0pt]{->}{PP}{VC}
		\ncline[nodesep=0pt]{->}{VC}{AC}
		\end{pspicture}
		\psset{unit=1.0mm}
        %\vspace*{0.2cm}
	\caption{A generic autonomous vehicle architecture \cite{LubinieckiEtAl:2020}.\label{Fig:Architecture}}
\end{figure}

Automated driving functions typically require a complex architecture, consisting of (at least) three main layers: Sensing, Planning, and Control. Each layer may be subdivided into multiple modules, as shown for the example architecture in Figure \ref{Fig:Architecture}. The Planning layer typically comprise of a module for \emph{decision making} and one or more modules for \emph{motion planning}, often moving gradually from global to local motion planning. The Control layer typically consists of a \emph{chassis controller} and individual \emph{actuator controllers}. 

The goal of motion planning is to create a reference trajectory (or path) that reflects a given decision on a specific driving goal, including when and where to drive, how to avoid obstacles, whether to change lanes, et cetera. The chassis controller transforms the trajectory into reference commands for all vehicle actuators, in particular, the steering system, the engine, and the brake. The objective is for the vehicle to track the given reference trajectory as closely as possible, while maintaining an intuitive, comfortable, and efficient behavior and obeying all actuator limits, as well as all dynamic and safety constraints.

An important drawback of a specialized architecture, like the one in Figure \ref{Fig:Architecture}, is the numerous interfaces between the different layers and modules, as each interface significantly increases the development efforts and poses a potential safety threat. Thus, it is desirable to reduce the number of required modules or, in other words, make each module as powerful as possible. 

In particular, for the interface between motion planning and chassis control, the reference trajectory should be drivable given the vehicle's dynamics and compatible with all relevant constraints. If these requirements are not (fully) satisfied by the motion planner, the controller may be difficult to tune, produce large tracking errors, or even become unstable. On the other hand, the planning of reference trajectories is a difficult problem that must be solved fast and with limited resources. Therefore, the planning task is often relaxed and simplified. Common approaches include decoupling of the lateral and longitudinal vehicle motion (i.e., computing paths instead of trajectories), neglecting dynamic feasibility (i.e., drivability of the path), ignoring actuator constraints, or sacrificing smoothness of the input signals \cite{DolgovEtAl:2010}. Other approaches choose a heuristic to find a coarse (sub-optimal) solution by limiting the search to a small subset of paths, e.g., defined by pre-selected basis functions or sampled at random \cite{WerlingEtAl:2010}.

In this context, Model Predictive Control (MPC) can be an attractive approach for chassis control. MPC is a holistic method to control nonlinear, multi-input multi-output (MIMO) systems with input and state constraints. As such, it is naturally suited for problems in vehicle control and has been used extensively in academic and industrial projects, e.g., \cite{DolgovEtAl:2010}, \cite{GaoEtAl:2010}, \cite{ZieglerEtAl:2014}, \cite{GutWer:2015}, \cite{KongEtAl:2015}, \cite{Lima:2016}. 

\subsection{Model Predictive Control (MPC)} 

Existing solvers for NMOPs are typically built on interior-point (IP) or sequential quadratic programming (SQP) methods \cite{DiehlEtAl:2009}. IP methods are highly effective for medium and large scale linear and nonlinear programs and they are known to have good robustness properties. Their most important drawback is a limited capability for \emph{warm-starting}, which may have a great impact on the computational performance of the MPC \cite{AxeHans:2008}. SQP methods solve a sequence of local quadratic programs (QPs), e.g., by an active set method. This is known to be efficient for MPC problems with moderate nonlinearities. For many driving scenarios, however, vehicle dynamics can be strongly nonlinear, in which case the convergence of SQP methods may be difficult or slow. Instead, the NAP algorithm used by \autompc\; performs a nonlinear line search, can be warm-started, and is guaranteed to preserve the primal feasibility of the solution, while monotonically improving the objective function value.

Many software packages are available for designing, simulating, and deploying MPC in the context of vehicle motion control, ranging from commercial toolboxes to open-source libraries. Particularly relevant examples are listed below:

\textbf{Model Predictive Control Toolbox} (\textit{MATLAB/Simulink\textregistered}): The MPC Toolbox is a comprehensive commercial toolbox that provides functions, simulation blocks and an app for developing MPC controllers. It supports linear (implicit, explicit, adaptive, gain-scheduled) and nonlinear MPC. It also offers deployable optimization solvers, based on C and CUDA code, and allows for the integration of custom solvers. Standard solvers include active-set, interior-point, and mixed-integer QP solvers. The MPC Toolbox is widely used in academia and industry due to its flexibility and integration with the MATLAB/Simulink\textregistered ecosystem.

\textbf{do-mpc} (\textit{Python}): do-mpc is a Python toolbox for Robust MPC of continuous and discrete-time nonlinear systems with uncertainty. It allows for defining system models, constraints, and objective functions, and then solves the optimal control problem using standard NLP solvers. For Robust MPC, it can employ multi-stage approaches that consider various scenarios of uncertain parameters.

\textbf{IPOPT} (\textit{FORTRAN/C++}): The Interior Point OPTimizer (IPOPT) is a widely used, open-source software package for solving large-scale NLP problems \cite{WaeBieg:2005}. It employs a primal-dual IP method, which is known for its robustness and efficiency in handling complex optimization problems with many variables and constraints. Due to generality of its NLP formulation, IPOPT has been used in a wide range of applications, including Linear and Nonlinear MPC.

\textbf{ACADO Toolkit} (\textit{C++/MATLAB}): The ACADO Toolkit is an open-source software framework for solving automatic control and dynamic optimization problems \cite{HouskaEtAl2:2011}. It can be used online and offline problems, however, it is primarily targeted for real-time Nonlinear MPC. It supports C++ code generation for fast, embedded application, employing variants of IP and SQP methods to solve the numerical optimization problem.

\textbf{GRAMPC} (\textit{C/C++/MATLAB/Python)}) GRAdient based MPC (GRAMPC) is an open-source software package specifically engineered to solve nonlinear optimal control problems on embedded hardware \cite{GraichenEtAl:2019}. It is focussed on systems with ultra-fast sampling rates in the (sub)millisecond range. GRAMPC builds on an augmented Lagrangian formulation combined with a gradient-based descent algorithm and implemented in C. This makes it particularly suitable for deployment on embedded hardware, including severely resource-constrained microcontrollers.

\textbf{PROPT} (\textit{MATLAB}): PROPT is a proprietary MATLAB-based software offered by Tomlab Optimization \cite{RutEd:2010}. It comes as an additional package for the TOMLAB Optimization Environment. PROPT is specifically designed for solving optimal control and parameter estimation problems that can be cast as continuous or mixed-integer NLP problems. It acts as a parser for TOMLAB's optimization solvers, including the Sparse Nonlinear OPTimizer (SNOPT) and Artelys' KNITRO, among additional external solvers that can be connected. PROPT provides extensive modeling capabilities for diverse constraint types, automatic detection of problem structures, and specialized support for non-smooth (hybrid) optimal control and problems involving binary or integer variables.

\textbf{FORCES PRO / PRODRIVER} (\textit{C/MATLAB}): FORCES PRO by Embotech is a commercial software framework for real-time embedded optimization, particularly multi-stage decision problems \cite{Domahidi:2013}. Instead of a pre-compiled library or a generic solver, FORCES PRO can be considered as a solver generation service. It provides tailored solver code for specific applications, including Linear and Nonlinear MPC, using algorithms such as IP and SQP methods. In addition to FORCES PRO, Embotech offers a dedicated software stack for automated driving, called PRODRIVER. It utilizes the customizable NLP solvers for autonomous navigation, motion planning, precise tracking, and robust control for single vehicles and vehicle fleets.

In summary, a variety of powerful software packages already exists for solving nonlinear, multi-stage optimization programs (NMOPs). Many of them have been successfully used for applications of vehicle motion control. Some of them can even generate stand-alone C code that is ready for rapid prototyping or deployable on ECUs.

\subsection{Novelty} 

The main novelty of the \autompc\; package is the implementation of a new nonlinear active set (NAS) algorithm in combination with a code generator, as an extension to prior work \cite{Schildi:2016}. It extends the optimization solver, which is usually provided by customary MPC packages, like the ones above, by a framework for vehicle path or trajectory tracking. This makes it ready to be used for vehicle control, yet highly configurable. Besides commercial products, such as Embotech's PRODRIVER, no comparable tool is currently available, in particular not as open source software or as a ROS package. 

The \autompc\; can be regarded as a monolithic function block for vehicle control. The main \emph{input} is a reference path or trajectory, besides the current state measurements, constraints, and tuning parameters. The reference path or trajectory can be of an arbitrary functional form (arc, clothoid, sigmoid, etc.). It essentially consists of a sequence of piecewise linear segments with additional reference parameters, such as the driving mode (forward, reverse, standstill). The main \emph{output} is the current control commands for the vehicle, including steering, engine, and brake. 

Since \autompc\; can handle low and high vehicle speeds and permits direction changes, it can be used for all kinds of driving scenarios, including highway, parking, etc. Safety constraints, such as obstacles or lane boundaries, are implemented via soft constraints. This guarantees the provision of a feasible control command by the MPC at all times. The most important features of \autompc\; can be summarized as follows:
\begin{itemize}
    \item robust and dependable MPC implementation, accounting for computational delays in motion planning, performing an automatic input checks and corrections, and including self-healing mechanisms in case of internal errors;
    \item generation of flat C code, with no external dependencies and static memory allocation, which is easy to compile and ready to be deployed on ECUs;
    \item designed as tested for the implementation on real vehicles and for rapid prototyping, e.g., based on dSpace\textregistered, Matlab/Simulink\textregistered or ROS;
    \item computationally efficient nonlinear optimization solver, with the possibility of warm starting;
    \item ready to be used with default settings and kinematic bicycle model, therefore quickly implemented even by non-experts; 
    \item the code is highly configurable in all essential components, such as the vehicle model, constraints, etc., therefore a flexible tool for vehicle control experts;
    \item suitable for most driving scenarios, including tracking of static or dynamics references (path or trajectory), forward and backward driving, mode switches, etc.;
    \item guaranteed feasibility of the MPC problem at all times, due to the use of soft state constraints;
    \item software package includes additional tools for visualization tools and several practical examples.
\end{itemize}

\subsection{Downloads}

The MPC code generator and the ROS package can be downloaded under the following links:\\
\hspace*{2.0cm}\href{https://git.ime.uni-luebeck.de/public-projects/asl/autompc}{\texttt{https://git.ime.uni-luebeck.de/public-projects/asl/autompc}}\\
\hspace*{2.0cm}\href{https://git.ime.uni-luebeck.de/public-projects/asl/ackermann_nlmpc}{\texttt{https://git.ime.uni-luebeck.de/public-projects/asl/ackermann\_nlmpc}}

\section{Model Predictive Control (MPC)}

As MPC involves numerical computations to find the control inputs, a fundamental closed-loop \emph{sampling rate} has to be defined by the user. In the case of \autompc, it is specified by a fundamental \emph{sampling time} $t_{\text{s}}>0$ (code parameter: \texttt{dt}).

\subsection{Model}

\autompc\; allows for the specification of an individual vehicle model, by means of a dedicated syntax in a separate text file. The model specification is automatically collected, compiled, and integrated into the final MPC code during code generation. 

\begin{inst}[frametitle=How to Specify the Vehicle Model]
    The vehicle model is specified via the vehicle model file, which must be named \texttt{mpcmodel.txt} and placed in the subfolder \texttt{mpcmodels} of the main folder. The first three lines of the vehicle model file define the model's
    \vspace*{-0.15cm}\begin{itemize}[leftmargin=.35cm,labelsep=.2cm]
        \item \texttt{states:} \textit{list of the state variables, separated by commas;}
        \item \texttt{inputs:} \textit{list of the control inputs, separated by commas;}
        \item \texttt{parameters:} \textit{(optional) list of model parameters, with a numerical value assigned by equal sign, separated by commas.}
    \end{itemize}\vspace*{-0.15cm}
    The first 5 states have to be: the global $X$-position (variable name \texttt{x}), the global $Y$-position (variable name \texttt{y}), the heading angle $\Phi$ (variable name \texttt{phi}), the velocity $v$ (variable name \texttt{v}), and the front wheel steering angle $\delta$ (variable name \texttt{delta}). The first 2 inputs have to be: the acceleration $a$ (variable name \texttt{a}), and the front wheel steering rate $\dot{\delta}$ (variable name \texttt{ddelta}). There are no mandatory parameters, so the list of parameters may be empty.\\
    The second part of the vehicle model file contains the equations of motion, i.e., a formula for the derivative of each state. The derivative is indicated by
    \begin{equation}\label{Equ:DefEoM}
        \texttt{dot(}\textit{state variable}\texttt{)=}\cdots\ef
    \end{equation}
    The right-hand side of equations \eqref{Equ:DefEoM} must represent a valid C-coded formula, which may contain basic functions defined in the \texttt{math.h} library as well as all of the states, inputs, and parameter variables, as defined in the first part of the vehicle model file. Each equation must be written in a separate line and ended with a semicolon.
\end{inst}

In principle, any vehicle model can be integrated. There is no restrictions on the number of states $n\geq 5$ or inputs $m\geq2$. Two examples of commonly used vehicle models are provided with the \autompc\; code package: a kinematic bicycle model (KBM) and a dynamic bicycle model (DBM) with linear tires \cite{KongEtAl:2015}. Additionally, a vehicle model file for a kinematic bicycle model with rear-wheel steering is provided, as an example with an additional control input ($m=3$).

\newpage
\renewcommand{\figurename}{\color{taucolorexam}Figure}
\begin{exam}[frametitle=Example: Kinematic Bicycle Model (KBM)]
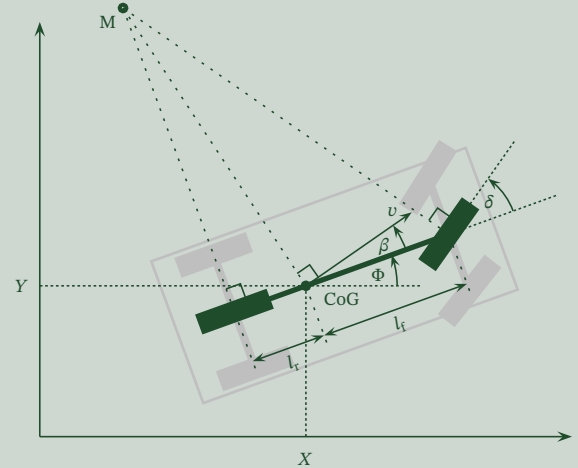
\begin{wrapfigure}{r}{0.42\textwidth}
	\footnotesize
    \begin{center}
		\begin{pspicture}(65,50)
		\rput{20}(35,20){
			% vehicle
			\psline[linewidth=8.0pt,linecolor=lightgray]{-}(-15,-8)(-5,-8) % rear tyre, right
			\psline[linewidth=8.0pt,linecolor=lightgray]{-}(-15,+8)(-5,+8) % rear tyre, left
			\psline[linewidth=3.0pt,linecolor=lightgray]{-}(-10,-8)(-10,+8)
			\rput{32}(20,-8){
				\psline[linewidth=8.0pt,linecolor=lightgray]{-}(-5,0)(+5,0) % front tyre, right
				}
			\rput{38}(20,+8){
				\psline[linewidth=8.0pt,linecolor=lightgray]{-}(-5,0)(+5,0) % front tyre, left
				}
			\psline[linewidth=3.0pt,linecolor=lightgray]{-}(20,-8)(20,+8)
			\pspolygon[linestyle=solid,linewidth=1.0pt,linecolor=lightgray](-18,-10)(-18,10)(26,10)(26,-10)(-18,-10)
			% bicycle
			\psline[linewidth=2.0pt,linecolor=taucolorexam]{-}(-10,0)(20,0)
			\psline[linewidth=8.0pt,linecolor=taucolorexam]{-}(-15,0)(-5,0) % rear tyre
			\rput{35}(20,0){
				\psline[linewidth=8.0pt,linecolor=taucolorexam]{-}(-5,0)(5,0) % front tyre
			}
			% CoG
			\pscircle[linewidth=2.5pt,linecolor=taucolorexam](0,0){0.1}
			\rput{-20}(3.8,-3.3){CoG}
			\psline[linewidth=0.6pt,linecolor=taucolorexam,linestyle=dashed,dash=1pt 3pt]{-}(-10,0)(-10,-8)
			\psline[linewidth=0.6pt,linecolor=taucolorexam,linestyle=dashed,dash=1pt 3pt]{-}(0,0)(0,-8)
			\psline[linewidth=0.6pt,linecolor=taucolorexam,linestyle=dashed,dash=1pt 3pt]{-}(20,0)(20,-8)
			\psline[linewidth=0.6pt,linecolor=taucolorexam,linestyle=solid,arrowsize=1.2]{<->}(-10,-7)(0,-7)
			\psline[linewidth=0.6pt,linecolor=taucolorexam,linestyle=solid,arrowsize=1.2]{<->}(0,-7)(20,-7)
			\rput[tc](-5,-9){$l_{\mathrm{r}}$}
			\rput[tc](10,-9){$l_{\mathrm{f}}$}
			% instantaneous center of rotation
			\pscircle[linewidth=2.5pt,linecolor=taucolorexam](-10,42.84){0.1}
			\rput{-20}(-12.5,41.84){M}
			\psline[linewidth=0.6pt,linecolor=taucolorexam,linestyle=dashed,dash=1pt 3pt]{-}(-10,0)(-10,42.84)
			\psline[linewidth=0.6pt,linecolor=taucolorexam,linestyle=dashed,dash=1pt 3pt]{-}(0,0)(-10,42.84)
			\psline[linewidth=0.6pt,linecolor=taucolorexam,linestyle=dashed,dash=1pt 3pt]{-}(20,0)(-10,42.84)
			\psline[linewidth=0.6pt,linecolor=taucolorexam]{-}(-10,3.2)(-8,3.2)(-8,1.2)
			\rput{15}(0,0){
				\psline[linewidth=0.6pt,linecolor=taucolorexam]{-}(0,2)(2,2)(2,0)
				}
			\rput{35}(20,0){
				\psline[linewidth=0.6pt,linecolor=taucolorexam]{-}(0,3.2)(2,3.2)(2,1.2)
				}
			% yaw angle
			\rput{-20}(0,0){
				\psline[linewidth=0.6pt,linecolor=taucolorexam,linestyle=dashed,dash=1pt 1pt]{-}(0,0)(15,0)
				\psarc[linewidth=0.6pt,linecolor=taucolorexam,linestyle=solid,arrowsize=1.2]{->}(0,0){12}{0}{20}
				\rput[mc](9.5,1.5){$\Phi$}
			}
			% steering angle	
			\psline[linewidth=0.6pt,linecolor=taucolorexam,linestyle=dashed,dash=1pt 1pt]{-}(20,0)(35,0)
			\rput{35}(20,0){
				\psline[linewidth=0.6pt,linecolor=taucolorexam,linestyle=dashed,dash=1pt 1pt]{-}(0,0)(15,0)
			}
			\psarc[linewidth=0.6pt,linecolor=taucolorexam,linestyle=solid,arrowsize=1.2]{->}(20,0){9}{0}{35}
			\rput{-20}(26.5,2.3){$\delta$}
			% sideslip angle
			\rput{15}(0,0){
				\psline[linewidth=0.6pt,linecolor=taucolorexam,linestyle=solid,arrowsize=1.2]{->}(0,0)(17,0)
				\rput{-35}(15,1.5){$v$}
			}
			\psarc[linewidth=0.6pt,linecolor=taucolorexam,linestyle=solid,arrowsize=1.2]{->}(0,0){14}{0}{15}
			\rput{-20}(11.5,1.4){$\beta$}
		}
		% coordinate system
		\psline[linewidth=0.6pt,linecolor=taucolorexam,arrowsize=4pt]{->}(0,0)(70,0)  % x
		\psline[linewidth=0.6pt,linecolor=taucolorexam,arrowsize=4pt]{->}(0,0)(0,55)  % y
		\psline[linewidth=0.6pt,linecolor=taucolorexam,linestyle=dashed,dash=1pt 1pt]{-}(35,20)(35,0)	
		\psline[linewidth=0.6pt,linecolor=taucolorexam,linestyle=dashed,dash=1pt 1pt]{-}(35,20)(0,20)
		\rput[tc](35,-3){$X$}
		\rput[mr](-1.5,20){$Y$}
		\end{pspicture}
	\end{center}
	\vspace*{0.1cm}
	\caption{\textcolor{taucolorexam}{Kinematic bicycle model (dark) of a four-wheel car (gray). CoG: center  of gravity, 
		O: instantaneous center of rotation.}
		\label{Fig:KinBiModel}}
	\vspace*{-0.1cm}
\end{wrapfigure}

\noindent The model assumptions of the KBM are illustrated in Figure \ref{Fig:KinBiModel}. For further details, the reader is referred to \cite{Rajamani:2012}. The equations of motion read as
\begin{equation}\label{Equ:Model}
\frac{\text{d}}{\text{d}t}
\begin{bmatrix}
X\\ Y\\ \Phi\\ v\\ \delta
\end{bmatrix} =
\begin{bmatrix}
v\,\cos\bigl(\Phi+\beta\bigr)\\ 
v\,\sin\bigl(\Phi+\beta\bigr)\\ 
v\,\cos\beta\,/\,\bigl(l_{f}+l_{r}\bigr)\,\tan\delta\\ 
a\\ 
\dot{\delta}
\end{bmatrix}\ec
\end{equation}
where
\begin{equation*}
\beta=\tan^{-1}\left(\frac{l_{r}}{l_{f}+l_{r}}\tan\bigl(\delta\bigr)\right)
\end{equation*}
is the sideslip angle. The expression $\frac{\text{d}}{\text{d}t}$ represents the derivative with respect to time $t$. The model \eqref{Equ:Model} has $n=5$ states (the global coordinates $X$, $Y$, $\Phi$ and the two augmented states $v$, $\delta$) and $m=2$ control inputs (the acceleration $a$ and the steering rate $\dot{\delta}$). The Center of Gravity (CoG) serves as the coordinate reference point of the vehicle. The vehicle model \eqref{Equ:Model} has two parameters, the distance from the CoG to the rear axle ($l_{r}$) and to the front axle ($l_{f}$); in this example, they are assumed to be given as
\begin{equation*}
    l_{r}=1.738\quad\text{and}\quad l_{f}=1.105\ef
\end{equation*}
For efficiency of the formulas, the two modified parameters are defined after an equivalent transformation,
\begin{equation*}
    l_{f}+l_{r}=2.843\quad\text{and}\quad \frac{l_{r}}{l_{f}+l_{r}}=0.6113\ef
\end{equation*}
The corresponding vehicle model file reads as shown in Code \ref{Cod:KBM}.
\begin{lstlisting}[caption=Example of the vehicle model file \texttt{KinematicBicycleModel.txt},label=Cod:KBM]
states: x, y, phi, v, delta
inputs: a, ddelta
parameters: l = 2.843 , lrlf = 0.6113

dot(x) = v * cos(phi + atan(lrlf*tan(delta)));
dot(y) = v * sin(phi + atan(lrlf*tan(delta)));
dot(phi) = v / l * cos(atan(lrlf*tan(delta))) * tan(delta);
dot(v) = a;
dot(delta) = ddelta;
\end{lstlisting}
\end{exam}

As explained above, the vehicle model is specified as an ordinary differential equation (ODE) in continuous time. It is converted to discrete time for numerical use inside MPC. The time steps $t_{k}=kt_{\text{s}}$ are denoted with $k=0,1,2,\dots,N$, based on the fundamental sampling time $t_{\text{s}}$ of the MPC. The state at time step $k$ is denoted $z_{k}\in\BR^{n}$ (vector including the states $X_{k}$, $Y_{k}$, $\Phi_{k}$, $v_{k}$, $\delta_{k}$, etc.). During each sample time period $k$, the input vector $u_{k}\in\BR^{m}$ (vector including the control inputs $a_{k}$, $\dot{\delta}_{k}$, etc.) is assumed to be constant. Given the state $z_{k}$ and an input $u_{k}$, the state at the next time step $z_{k+1}$ can be computed based on some method for numerical integration, which can be manually selected in \autompc.

\begin{inst}[frametitle=How to Select the Integration Method]
\noindent The selected integration method for solving the ODE is hardcoded in the final MPC code. It is specified by the variable \texttt{intmethod} in the code generation file. 
\begin{description}
    \item[\texttt{intmethod = 1}:] Runge-Kutta 1 explicit (Euler explicit)
    \item[\texttt{intmethod = 2}:] Runge-Kutta 2 explicit (midpoint method)
    \item[\texttt{intmethod = 3}:] Runge-Kutta 3 explicit (Simpson's rule)
    \item[\texttt{intmethod = 4}:] Runge-Kutta 3 explicit (Heun's method)
    \item[\texttt{intmethod = 5}:] Runge-Kutta 4 explicit (classical RK)
    \item[\texttt{intmethod = 6}:] Runge-Kutta 1 implicit (Euler implicit)
    \item[\texttt{intmethod = 7}:] Runge-Kutta 2 implicit (trapezoidal rule)
\end{description}
The explicit Runge-Kutta methods become more accurate with increasing order, but also more computationally demanding \cite{Butcher:2008}. The implicit methods are even more computationally expensive, but may be more suitable for particular vehicle models and application settings \cite{Butcher:2008}, as they involve the solution to a nonlinear equation system using the Newton-Raphson method. A solution tolerance and maximum number of iterations for the Newton-Raphson method are defined by the parameters \texttt{newtontol} and \texttt{newtonit} in the code generation file (default: \texttt{newtontol = 1e-14} and \texttt{newtonit = 10}).\\
The accuracy of the integration can also be increased, independently of choosing a higher-order integration method. A finer mesh for the numerical integration---compared to the fundamental sampling time $t_{\mathrm{s}}$ of the MPC---can be selected by including additional support nodes within each fundamental sampling interval. The number of support nodes $k_{\mathrm{supnds}}$ is specified via the parameter \texttt{supnds}, which must be a non-negative integral number (default: \texttt{supnds = 0}). The support nodes are evenly spaced, thus leading to an integration step length of $t_{\mathrm{s}}/(1+k_{\mathrm{supnds}})$.
\end{inst}

Corresponding to the selected integration scheme, the discretized vehicle dynamics are denoted
\begin{equation}\label{Equ:DiscreteModel}
  z_{k+1}=f\bigl(z_{k},u_{k}\bigr)\ec\quad\text{for}\;k=0,1,\dots,N-1\ec
\end{equation}
where $f:\BR^{n\times m}\rightarrow\BR^{n}$ is an implicitly defined function that is repeatedly evaluated numerically during the execution of the \texttt{gsnlmpc} code.

\textbf{Notation.} The concept of MPC is to solve a Finite-Time Optimal Control Problem (FTOCP) in a receding horizon fashion, based on predictions by the system model \eqref{Equ:DiscreteModel}, and to apply only the first control input of the optimal solution \cite{BoBeMo:2017,RaMaDi:2018}. For notational simplicity, the following presentation is limited to a single time step, $k=0$. Hence the optimal control inputs to be solved for are denoted $u_{0},u_{1},\dots,u_{N-1}$. The corresponding (model-predicted) \emph{open-loop states} of the system are denoted $z_{0},z_{1},\dots,z_{N}$, where the initial state $z_{0}$ is given by a measurement, or estimate $\hat{z}_{0}$, of the true state of the vehicle.

\subsection{Reference}
\label{Chap:Reference}

\begin{figure}[h]
\psset{unit=0.9mm}
\begin{center}
\begin{pspicture}(0,-10)(120,44)
	\scriptsize
	\bol
	\psset{arrowsize=7pt}
    % nodes
    \psline[linecolor=MedBrown]{-*}(100,12)(112,16)
    \rput[tl](111,14.5){\textcolor{MedBrown}{$\displaystyle\begin{bmatrix}x_{N_{\mathrm{p}}}\\ y_{N_{\mathrm{p}}}\end{bmatrix}$}}
    \psline[linecolor=MedViolet]{-*}(86,10)(100,12)
    \rput[tl](93,10.5){\textcolor{MedViolet}{$\displaystyle\begin{bmatrix}x_{N_{\mathrm{p}}-1}\\ y_{N_{\mathrm{p}}-1}\end{bmatrix}$}}
    \psline[linecolor=MedGray,linestyle=dashed,dash=2pt 2pt]{-*}(50,17)(86,10)
    \psline[linecolor=MedYellow]{-*}(36,18)(50,17)
    \rput[tl](49,15.5){\textcolor{MedYellow}{$\displaystyle\begin{bmatrix}x_{4}\\ y_{4}\end{bmatrix}$}}
    \psline[linecolor=MedGreen]{-*}(22,16)(36,18)
    \rput[tl](35,16.5){\textcolor{MedGreen}{$\displaystyle\begin{bmatrix}x_{3}\\ y_{3}\end{bmatrix}$}}
    \psline[linecolor=MedBlue]{-*}(10,10)(22,16)
    \rput[tl](21,14.5){\textcolor{MedBlue}{$\displaystyle\begin{bmatrix}x_{2}\\ y_{2}\end{bmatrix}$}}
    \psline[linecolor=MedRed]{-*}(0,0)(10,10)
    \rput[tl](9,8.5){\textcolor{MedRed}{$\displaystyle\begin{bmatrix}x_{1}\\ y_{1}\end{bmatrix}$}}
    \psline[linecolor=black]{-*}(0,0)(0,0)
    \rput[tl](-1,-1){$\displaystyle\begin{bmatrix}x_{0}\\ y_{0}\end{bmatrix}$}
    % segments
    \rput[br](6,6){\textcolor{MedRed}{$\displaystyle\begin{bmatrix}\varphi_{1}\\ v_{1}\\ \delta_{1}\\ l_{1}\end{bmatrix}$}}
    \rput[br](17,13.5){\textcolor{MedBlue}{$\displaystyle\begin{bmatrix}\varphi_{2}\\ v_{2}\\ \delta_{2}\\ l_{2}\end{bmatrix}$}}
    \rput[br](30,17.5){\textcolor{MedGreen}{$\displaystyle\begin{bmatrix}\varphi_{3}\\ v_{3}\\ \delta_{3}\\ l_{3}\end{bmatrix}$}}
    \rput[bl](42,18){\textcolor{MedYellow}{$\displaystyle\begin{bmatrix}\varphi_{4}\\ v_{4}\\ \delta_{4}\\ l_{4}\end{bmatrix}$}}
    \rput[br](95,12){\textcolor{MedViolet}{$\displaystyle\begin{bmatrix}\varphi_{N_{\mathrm{p}}-1}\\ v_{N_{\mathrm{p}}-1}\\ \delta_{N_{\mathrm{p}}-1}\\ l_{N_{\mathrm{p}}-1}\end{bmatrix}$}}
    \rput[br](108,15){\textcolor{MedBrown}{$\displaystyle\begin{bmatrix}\varphi_{N_{\mathrm{p}}}\\ v_{N_{\mathrm{p}}}\\ \delta_{N_{\mathrm{p}}}\\ l_{N_{\mathrm{p}}}\end{bmatrix}$}}
\end{pspicture}
\end{center}
\vspace*{-0.3cm}
\caption{Illustration of the MPC reference, path or trajectory.\label{Fig:Reference}}
\end{figure}
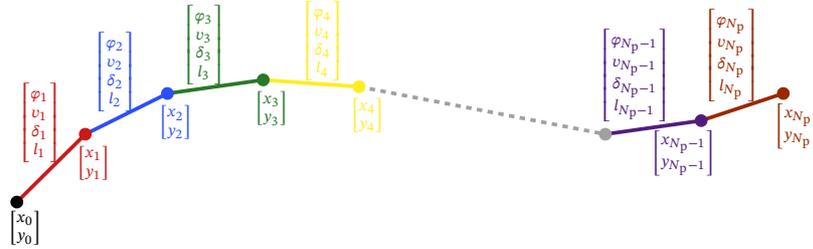

The \emph{reference} is the main input to the MPC during runtime, besides the state measurements. The reference is provided in the form of either a \emph{reference path} or a \emph{reference trajectory}. Both consist of a piecewise linear, planar curve, which is additionally parameterized, as illustrated in Figure \ref{Fig:Reference}. The piecewise linear elements are referred to as \emph{segments}.

\renewcommand{\figurename}{\color{black}Figure}

The key difference between a path and a trajectory is that the latter also carries information about timing, and hence includes a simultaneous level of position control in the MPC. More specifically, if the vehicle is lagging behind on a \emph{trajectory}, it will deliberately accelerate (beyond the reference speed) in order to catch up; conversely, if it is moving ahead, it will deliberately decelerate in order to synchronize with the time schedule along the trajectory. In contrast, when tracking a \emph{path}, the goal is only to follow a geometric curve with the given reference velocity, independently of time. However, the unifying term `\emph{trajectory}' is occasionally used in the following, instead of `path or trajectory', because the definition of a path and a trajectory are the same, only the timing information is ignored.

The code further distinguishes two types of reference paths, a regular path and a circular path. A \emph{circular path} is supposed to be tracked repeatedly, such as a lap on a race track. When reaching the end of the circular path, the MPC automatically continues restarts the path from the beginning. The shape of the circular path is arbitrary, i.e., it does not have to be a circle. However, the end point should be close to (or identical with) the starting point. In case of a \emph{regular path}, or a \emph{trajectory}, if the vehicle reaches the end of the final segment, the MPC performs an emergency stop. Hence, a new path or trajectory should always be provided well before the vehicle finishes the current one.

A new reference trajectory is provided by simply increasing the time stamp $T$ of the trajectory. As long as the time stamp is lower or remains the same as the previous one, \autompc\; works with the 

There are few restrictions on the geometric shape of the trajectory. In particular, it does not have to be drivable or of a specific functional form (arc, clothoid, sigmoid, etc.). It may cross itself and it may even pass through obstacles. In fact, obstacles can be avoided by the additional specification of state constraints, as discussed further below. Any path or trajectory may also contain intermediary stops and/or direction changes.

\newpage
\renewcommand{\figurename}{\color{taucolorinst}Figure}
\begin{inst}[frametitle=How to Define the Reference Path / Trajectory]
\begin{wrapfigure}{r}{0.32\textwidth}
    \footnotesize
    \vspace*{-0.3cm}
    \begin{center}
    \psset{unit=0.7mm}
    
    \begin{pspicture}(-40,-40)(40,40)
            % Global coordinates
			\psline[linecolor=taucolorinst,linewidth=0.8pt,arrowsize=5pt]{->}(-40,-10)(40,-10)
			%\rput[tr](38,-11.5){$x$}
			\psline[linecolor=taucolorinst,linewidth=0.8pt,arrowsize=5pt]{->}(-10,-40)(-10,40)
			%\rput[tr](-11.5,38){$y$}
			% Local coordinates
			\pscircle[linecolor=taucolorinst,linewidth=0.1pt,fillstyle=solid,fillcolor=taucolorinst](20,12){0.4}
			\rput{30}(20,12){
			\psline[linecolor=taucolorinst,linewidth=0.8pt,arrowsize=5pt]{->}(-20,0)(20,0)
			\rput[tr](18,-2){$x$}
			\psline[linecolor=taucolorinst,linewidth=0.8pt,arrowsize=5pt]{->}(0,-20)(0,20)
			\rput[tr](-2,18){$y$}
			}
			% Dashed lines
			\psline[linecolor=taucolorinst,linewidth=0.8pt,linestyle=dashed,dash=2pt 2pt]{-}(20,-10)(20,12)
			\psline[linecolor=taucolorinst,linewidth=0.8pt,linestyle=dashed,dash=2pt 2pt]{-}(-10,12)(32,12)
			\psline[linecolor=taucolorinst,linewidth=0.8pt]{-}(20,-10.5)(20,-9.5)
			\rput[tc](20,-13){$X$}
			\psline[linecolor=taucolorinst,linewidth=0.8pt]{-}(-10.5,12)(-9.5,12)
			\rput[tc](-13,12){$Y$}
			% Rotation
			\psarc[linecolor=taucolorinst,linewidth=0.8pt,arrowsize=5pt]{->}(20,12){10}{0}{30}
			\rput[mc](27,13.5){$\Phi$}
			% Point C
			%\pscircle[linewidth=0.1pt,fillstyle=solid,fillcolor=black](27,28){0.4}
			%\rput[tr](26,27){$C$}
        \end{pspicture}
      \end{center}
      \caption{\textcolor{taucolorinst}{Illustration of global coordinate system $(X,Y)$ and local coordinate system $(x,y)$ used for the reference path / trajectory.}\label{Fig:GlobalLocalCoordinates}}
  \end{wrapfigure}
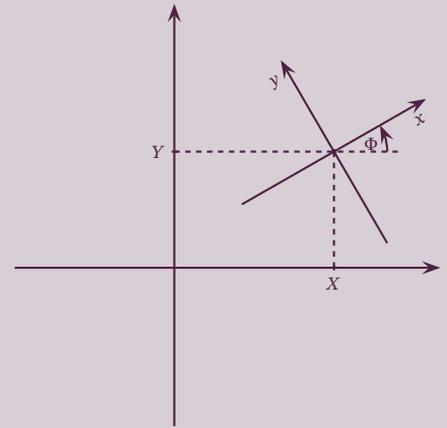
  
  \noindent The reference trajectory (or path) consists of a \emph{trajectory header} and a \emph{trajectory body}. The trajectory header contains global information about the trajectory, the trajectory body contains a specification of the segments. The entries of the trajectory header are listed and explained in Table \ref{Tab:TrajHeader}.
  
  The \emph{root} of the trajectory $(X,Y)$ is the starting point of the trajectory and, simultaneously, defines the origin of the local coordinate system, as shown in Figure \ref{Fig:GlobalLocalCoordinates}. Moreover, the local coordinate system is rotated by the orientation angle $\Phi$.
  
  The \emph{global time (stamp)} $T$ of the reference trajectory corresponds to the time when the vehicle is supposed to be at the starting point and, simultaneously, represents its \emph{time stamp}. The MPC always uses the reference trajectory with the latest time stamp. To this end, new reference trajectories may be provided at any time, i.e., not necessarily with a regular frequency. However, a new reference trajectory should be provided on time before the vehicle reaches the end of the current reference trajectory, as otherwise the MPC triggers a braking maneuver to stop the vehicle.

  The variable $P_{\mathrm{type}}\in\{0,1,2\}$ defines the \emph{reference type}, i.e., either a trajectory ($P_{\mathrm{type}}=0$), a regular path ($P_{\mathrm{type}}=1$), or a circular path ($P_{\mathrm{type}}=2$). The number $S\in\BN$ defines the number of active segments in the reference trajectory, which does not need to be constant over time. There must always be at least one segment, $S\geq 1$. Due to static memory allocation throughout the code, there is an upper limit to the number of segments, $S\leq S_{\max}$, where $S_{\max}$ is a parameter (named \texttt{Nn}) that may be chosen during code generation.

  \begin{center}
    \begin{tabular}{|p{0.07\textwidth}|p{0.07\textwidth}|p{0.07\textwidth}|p{0.4\textwidth}|}
    \hline
    \textbf{Name} & \textbf{Type} & \textbf{Entry} & \textbf{Description}\\\hline\hline
    $T$ & \texttt{double} & \hspace*{0.55cm}0 & global time (stamp) of the trajectory [s]\\\hline
    $X$ & \texttt{double} & \hspace*{0.55cm}1 & global $X$-position of the root of the trajectory [m]\\\hline
    $Y$ & \texttt{double} & \hspace*{0.55cm}2 & global $Y$-position of the root of the trajectory [m]\\\hline
    $\Phi$ & \texttt{double} & \hspace*{0.55cm}3 & rotation of the local coordinate system [rad]\\\hline
    $P_{\mathrm{type}}$ & \texttt{uint8} & \hspace*{0.55cm}4 & trajectory type: \texttt{0} = trajectory, \texttt{1} = regular path, \texttt{2} = circular path\\\hline
    $S$ & \texttt{int64} & \hspace*{0.55cm}5 & number of active segments in the trajectory, where $1\leq S\leq S_{\max}$\\\hline
    \end{tabular}\\\vspace*{0.4cm}
     \small
     \textbf{\textsf{Table \refstepcounter{table}\label{Tab:TrajHeader}\thetable.}} Entries of the trajectory header.
  \end{center}\vspace*{0.2cm}

  The trajectory body forms the main part of the reference trajectory. It is represented by $S$ linear segments, as illustrated in Figure \ref{Fig:Reference}. Each segment, numbered as $i=1,2,\dots,S$, consists of a line with an end point, called a \emph{(trajectory) node}. The starting point of each segment corresponds to the end point of the previous segment, except for the first segment, which starts at the root of the trajectory. 
  
  Table \ref{Tab:TrajBody} provides an overview of the data used to specify \emph{each} segment. The segments are listed consecutively, i.e., the reference trajectory consists of $(6+11\,S)$ (meaningful) numbers: $6$ for the trajectory header and $11$ for each of the $S$ trajectory nodes. The remaining memory, towards the $(6+11\,S_{\max})$ numbers in total, which are allocated for the interface of the reference trajectory, will be ignored and may thus be filled with arbitrary values.

  \begin{center}
    \begin{tabular}{|p{0.07\textwidth}|p{0.07\textwidth}|p{0.07\textwidth}|p{0.4\textwidth}|}
    \hline
    \textbf{Name} & \textbf{Type} & \textbf{Entry} & \textbf{Description}\\\hline\hline
    $t^{(i)}$ & \texttt{double} & \hspace*{0.55cm}0 & local time for passing the segment node [s]\\\hline
    $x^{(i)}$ & \texttt{double} & \hspace*{0.55cm}1 & local $x$-position of the segment node [m]\\\hline
    $y^{(i)}$ & \texttt{double} & \hspace*{0.55cm}2 & local $y$-position of the segment node [m]\\\hline
    $\varphi^{(i)}$ & \texttt{double} & \hspace*{0.55cm}3 & angle between the segment and the $x$-axis of the local coordinate system [rad]\\\hline
    $v^{(i)}$ & \texttt{double} & \hspace*{0.55cm}4 & reference velocity along the segment, $v^{(i)}\geq 0$, as the driving direction is controlled by the driving mode [m/s]\\\hline
    $a^{(i)}$ & \texttt{double} & \hspace*{0.55cm}5 & reference acceleration along the segment [m/s$^2$]\\\hline
    $\delta^{(i)}$ & \texttt{double} & \hspace*{0.55cm}6 & reference front axle steering angle along the segment [rad]\\\hline
    $\beta^{(i)}$ & \texttt{double} & \hspace*{0.55cm}7 & reference vehicle sideslip angle along the segment [rad]\\\hline
    $D^{(i)}$ & \texttt{uint8} & \hspace*{0.55cm}8 & driving mode along the segment: \texttt{0} = standstill, \texttt{1} = forward, \texttt{2} = reverse\\\hline
    $d_{\mathrm{left}}^{(i)}$ & \texttt{double} & \hspace*{0.55cm}9 & state constraint, maximum lateral deviation to the left of the segment [m]\\\hline
    $d_{\mathrm{right}}^{(i)}$ & \texttt{double} & \hspace*{0.55cm}10 & state constraint, maximum lateral deviation to the right of the segment [m]\\\hline
    \end{tabular}\\\vspace*{0.4cm}
     \small
     \textbf{\textsf{Table \refstepcounter{table}\label{Tab:TrajBody}\thetable.}} Entries of each segment $i\in\{1,2,\dots,S\}$ of the trajectory body.
  \end{center}\vspace*{0.2cm}

  The segment node $i$, i.e., the endpoint of the segment $i$, is specified by $x^{(i)}$ and $y^{(i)}$ in local coordinates. Note that the position data contained in the state measurements $z_{k}$ are, nonetheless, always in global coordinates. The reference time $t^{(i)}$ refers to the time for the vehicle to pass the segment node $i$. This timing information is irrelevant in the case of a reference path ($P_{\mathrm{type}}^{(i)}=1,2$). The value of $\varphi^{(i)}$ specifies the angle of the segment with the $x$-axis of the local coordinate system. As the angle is implicitly defined by the segment nodes $i$ and $i-1$, the angle is redundant information; it is used to avoid unnecessary and mirrored re-calculations inside the MPC. 

  \begin{wrapfigure}{r}{0.4\textwidth}
	\footnotesize
	\vspace*{-0.5cm}
	\begin{center}
		\begin{pspicture}(87,45)
		% coordinate system
		\psline[linecolor=taucolorinst,linewidth=0.6pt,arrowsize=4pt]{->}(0,0)(70,0)  % x
		\psline[linecolor=taucolorinst,linewidth=0.6pt,arrowsize=4pt]{->}(0,0)(0,45)  % y
		\rput[tc](70,-3){$x$}
		\rput[mr](-1.5,45){$y$}
		% obstacle
		\pspolygon[linecolor=taucolorinst,fillcolor=taucolorinst,linestyle=solid,linewidth=1.0pt,fillstyle=hlines,
		hatchwidth=0.2pt,hatchsep=5pt,hatchangle=40](37,15)(37,25)(53,25)(53,15)
		% reference path
		\psline[linecolor=taucolorinst,linewidth=1.4pt,arrowsize=7pt]{-*}(33,20)(57,20)
		\psline[linecolor=taucolorinst,linewidth=1.4pt,arrowsize=7pt]{-*}(18,17)(33,20)
		\psline[linecolor=taucolorinst,linewidth=1.4pt,arrowsize=7pt]{*-*}(6,12)(18,17)
		% right constraints
		\psline[linecolor=taucolorinst,linewidth=1.0pt,linestyle=dashed,dash=2pt 2pt]{-}(33,26)(57,26)
		\psline[linecolor=taucolorinst,linewidth=1.0pt,linestyle=dashed,dash=2pt 2pt]{-}(15,29)(30,32)
		\psline[linecolor=taucolorinst,linewidth=1.0pt,linestyle=dashed,dash=2pt 2pt]{-}(9,4)(20,9)
		% left constraints
		\psline[linecolor=taucolorinst,linewidth=1.0pt,arrowsize=4pt,linestyle=dashed,dash=2pt 2pt]{-}(33,40)(57,40)
		\psline[linecolor=taucolorinst,linewidth=1.0pt,linestyle=dashed,dash=2pt 2pt]{-}(18.5,15)(33.5,18)
		\psline[linecolor=taucolorinst,linewidth=1.0pt,linestyle=dashed,dash=2pt 2pt]{-}(3,20)(16,25)
		% text
        \psline[linecolor=taucolorinst,linewidth=0.9pt,arrowsize=4pt]{->}(12,14.5)(14.8,6.5)
        \psline[linecolor=taucolorinst,linewidth=0.9pt,arrowsize=4pt]{->}(12,14.5)(9.2,22.5)
		\rput[tr](13.5,10.5){$d_{\mathrm{right}}^{(1)}>0$}
		\rput[br](10.5,15.5){$d_{\mathrm{left}}^{(1)}>0$}
        \psline[linecolor=taucolorinst,linewidth=0.9pt,arrowsize=4pt]{->}(25.5,18.5)(25.96,16.2)
        \psline[linecolor=taucolorinst,linewidth=0.9pt,arrowsize=4pt]{->}(25.5,18.5)(23,31)
		\rput[tr](29.5,15.5){$d_{\mathrm{right}}^{(2)}>0$}
		\rput[br](24,20.5){$d_{\mathrm{left}}^{(2)}>0$}
        \psline[linecolor=taucolorinst,linewidth=0.9pt,arrowsize=4pt]{->}(35,20)(35,26)
        \psline[linecolor=taucolorinst,linewidth=0.9pt,arrowsize=4pt]{->}(36,20)(36,40)
		\rput[tr](35,29.5){$d_{\mathrm{right}}^{(3)}<0$}
		\rput[br](35.5,35){$d_{\mathrm{left}}^{(3)}>0$}

		\end{pspicture}
	\end{center}
	\vspace*{-0.1cm}
	\caption{\textcolor{taucolorinst}{Example of reference path (solid line) with three segments, using state constraints (dashed lines) to avoid a rectangular obstacle (hatched gray area).}
	\label{Fig:StateConstraints}}
	\vspace*{-0.1cm}
  \end{wrapfigure}
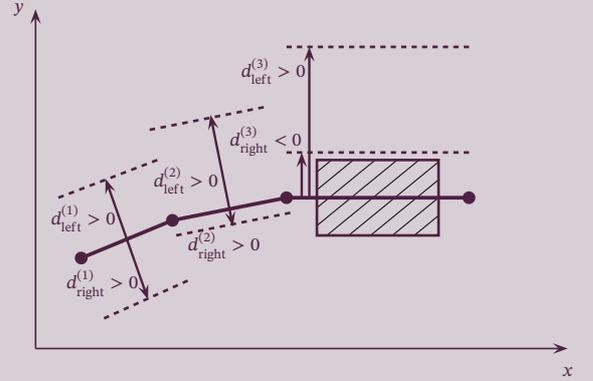
  
  The reference velocity and acceleration, $v^{(i)}$ and $a^{(i)}$, are also, in a sense, redundant, when moving along the segments of the trajectory. It is recommended to choose them somewhat consistently along the trajectory, even though this is not a strict requirement. The reference velocity must always be positive, $v^{(i)}\geq 0$. In fact, direction changes are caused by switching the reference driving mode $D^{(i)}$ between two segments. It is advisable to include a standstill segement ($D^{(i)}=0$) in between a change of direction from forward ($D^{(i-1)}=1$) to reverse ($D^{(i+1)}=2$), or vice versa. This leaves time, e.g., for a static steering maneuver or for a necessary gear shift. The values of $\delta^{(i)}$ and $\beta^{(i)}$ provide references for the front axle steering angle and the vehicle sideslip angle along the segment, respectively. Note that the sideslip angle reference is ignored in the current version of the code. The references for further states are simply assumed as zero.

  The final two values, $d_{\mathrm{left}}^{(i)}$ and $d_{\mathrm{right}}^{(i)}$, of the trajectory body represent the state constraints of the respective segment. They refer to the distance to a left and right boundary parallel to the segment, defining a safe corridor; see Figure \ref{Fig:StateConstraints} for an illustration. The attributes `left' and `right' refer to the direction of moving along the trajectory. In particular, they do not refer the vehicle's actual orientation, e.g., when driving in reverse. Each of the distances may also be negative, as it is the case for $d_{\mathrm{right}}^{(3)}$ in the example of Figure \ref{Fig:StateConstraints}. In other words, the reference trajectory may actually pass through obstacles, whence the actual collision avoidance is a task of the MPC.
\end{inst}

\textbf{Reference Generator.} The reference path or trajectory is translated by a \emph{reference generator} inside the \autompc\; code into specific reference values for the inputs and states over the prediction horizon. These reference values are denoted $u^{\mathrm{(ref)}}_{k}\in\BR^{m}$ and $z^{\mathrm{(ref)}}_{k}\in\BR^{n}$ in vector form, respectively, and 
\begin{equation}\label{Equ:ReferenceInputsStates}
    a^{\mathrm{(ref)}}_{k}\ec\quad \gamma^{\mathrm{(ref)}}_{k}\ec\quad
    x^{\mathrm{(ref)}}_{k}\ec\quad y^{\mathrm{(ref)}}_{k}\ec\quad \varphi^{\mathrm{(ref)}}_{k}\ec\quad v^{\mathrm{(ref)}}_{k}\ec\quad\delta^{\mathrm{(ref)}}_{k}\qquad\fa k=1,2,\dots,N\ec
\end{equation}
for the individual inputs and states. Note that \eqref{Equ:ReferenceInputsStates} lists only the references for the two mandatory inputs and for the five mandatory states. If the MPC model has further inputs ($m>2$) and/or states ($n>5$), the corresponding reference values are set to zero.

\textbf{Objective Function.} The objective of the FTOCP is to minimize the weighted sum of the quadratic deviations of the planned control inputs $u_{0},u_{1},\dots,u_{N-1}$ and the model-predicted states $z_{1},z_{2},\dots,z_{N}$ to the reference values. Specifically, the \emph{cost function} consists of $N$ \emph{input stage cost} terms $\ell_{\mathrm{u}}:\BR^{m}\to\BR^{+}$,
\begin{equation}\label{Equ:InputCostFunction}
	\ell_{\mathrm{u}}\bigl(u_{k})\triangleq
	r_{1}\Bigl(a_{k}-a^{\mathrm{(ref)}}_{k}\Bigr)^2+
    r_{2}\Bigl(\gamma_{k}-\gamma^{\mathrm{(ref)}}_{k}\Bigr)^2+
    \sum_{j=3}^{m}
	r_{j}u_{k}(j)^2
\end{equation}
for $k=0,1,\dots,N-1$, and $N$ \emph{state stage cost} terms
$\ell_{\mathrm{z}}:\BR^{n}\to\BR^{+}$,
\begin{equation}\label{Equ:StateCostFunction}
	\ell_{\mathrm{z}}\bigl(z_{k})\triangleq
	q_{1}\Bigl(s_{k}-s^{\mathrm{(ref)}}_{k}\Bigr)^2+
    q_{2}l_{k}^2+
    q_{3}\Bigl(\varphi_{k}-\varphi^{\mathrm{(ref)}}_{k}\Bigr)^2+
    q_{4}\Bigl(v_{k}-v^{\mathrm{(ref)}}_{k}\Bigr)^2+
    q_{5}\Bigl(\delta_{k}-\delta^{\mathrm{(ref)}}_{k}\Bigr)^2+
    \sum_{j=6}^{n}
	q_{j}z_{k}(j)^2
\end{equation}
for $k=1,2,\dots,N$. Here $u_{k}(j)$ and $z_{k}(j)$ denote the $j$-th component of the vectors $u_{k}$ and $z_{k}$, respectively. The scalars $r_{1},r_{2},\dots,r_{m}>0$ constitute the (positive) \emph{input weights} and the scalars $q_{1},q_{2},\dots,q_{n}\geq 0$ constitute the (non-negative) \emph{state weights}.

\textbf{Localization.} The generation of the reference values \eqref{Equ:ReferenceInputsStates} is based on the \emph{localization} of the vehicle on the reference trajectory. The main principle for this localization is a projection of the vehicle position onto the spatial trajectory; in other words, the \emph{localization point} is defined as the closest point on the reference trajectory to the current vehicle position. However, there are some caveats to this principle. First, in the case of self-intersecting reference trajectories, it is not desirable to jump, either back or ahead, to a different part of the trajectory. Second, the search should be restricted to those parts of the trajectory with the same driving mode as the current driving mode of the vehicle. Third, to reduce the computational effort, it is not reasonable to search the full reference trajectory for the closest point to the vehicle, since a good initial guess is available from the previous time step.

Therefore, the search algorithm retracts a certain number of segments, $S_{\mathrm{search}}>0$ (code parameter: \texttt{segsearch}), from the trajectory segment that contains the previous localization point. The search then proceeds forward, segment by segment, by calculating the projection point for each segment, and the corresponding distance to the vehicle. The goal is to find a local minimum in this distance. To this end, the search terminates if no new minimum has been found for the last $S_{\mathrm{search}}$ segments.

An appropriate value of $S_{\mathrm{search}}$ depends on the typical length of the segments in the reference trajectory. If a high number of short segments is used, i.e., a high resolution of the trajectory, then $S_{\mathrm{search}}$ should be chosen higher; to improve computational efficiency, $S_{\mathrm{search}}$ should be chosen lower. Choosing $S_{\mathrm{search}}=S_{\mathrm{max}}$ means that the search always proceeds over the entire reference trajectory (which may cause problems in the case of self-intersecting trajectories).

\textbf{Additional parameters for reference trajectories.} For the particular case of a reference trajectory (i.e., $P_{\mathrm{type}}=0$), two additional parameters are available that affect the reference generation. In fact, for a trajectory, the reference velocity is slightly increased (decreased) from the original segment values if the vehicle is lagging behind (is getting ahead). The degree of modification of the reference velocity can be quantified by the so-called \emph{catch-up time} (code parameter: \texttt{cuptime}). It indicates the period of time in which the vehicle is supposed to catch up with the timed reference position on the trajectory (in seconds). Moreover, the maximum relative modification of the reference velocity can be bounded, by an additional parameter called \texttt{maxrefvelmod}.

\subsection{Input and State Constraints}

\textbf{Input Constraints.} The \autompc\; code provides for individual, constant actuator bounds and rate constraints over the prediction horizon. In particular, the \emph{input bounds} are defined as
\begin{subequations}\label{Equ:InputBounds}\begin{align}
	a_{\min}     \leq &a_{k}      \leq a_{\max}\ec\\
	\gamma_{\min}\leq &\gamma_{k} \leq \gamma_{\max}\ec\\
    u_{\min}(j)\leq   u_{k}&(j)    \leq u_{\max}(j)\qquad\fa j=3,\dots,m\ec
\end{align}\end{subequations}
for all $k=0,\dots,N-1$, and the \emph{input rate constraints} are defined as
\begin{subequations}\label{Equ:InputRates}\begin{align}
	\Delta a_{\min}     \leq \frac{1}{t_{\mathrm{s}}}\Bigl(a_{k}&-a_{k-1}\Bigr)      \leq \Delta a_{\max}\ec\\
	\Delta \gamma_{\min}\leq \frac{1}{t_{\mathrm{s}}}\Bigl(\gamma_{k}&-\gamma_{k-1}\Bigr) \leq \Delta \gamma_{\max}\ec\\
    \Delta u_{\min}(j)\leq   \frac{1}{t_{\mathrm{s}}}\Bigl(u_{k}(j)&-u_{k-1}(j)\Bigr)    \leq \Delta u_{\max}(j)\qquad\fa j=3,\dots,m\ec
\end{align}\end{subequations}
for all $k=0,\dots,N-1$. Note that $\Delta a_{\min}$, $\Delta \gamma_{\min}$, $\Delta u_{\min}(j)$ and $\Delta a_{\max}$, $\Delta \gamma_{\max}$, $\Delta u_{\max}(j)$ represent the minimum and maximum derivatives of the longitudinal acceleration, the steering rate, and the remaining inputs $j$, respectively. In \eqref{Equ:InputRates} for $k=0$, the values of $a_{-1}$, $\delta_{-1}$ and $u_{-1}(j)$ are given by the corresponding control inputs that have been applied to the vehicle in the time step previous to $k=0$.\footnote{Recall that the presentation refers to the setup of the FTOCP in time step $k=0$, where only the first optimal control inputs $a_{0}\opt$, $\delta_{0}\opt$ and $u_{0}\opt(j)$ are applied to the vehicle. Thus $a_{-1}$, $\delta_{-1}$ and $u_{-1}(j)$ equal to the first optimal control inputs from the FTOCP in the time step previous to $k=0$.}

\renewcommand{\figurename}{\color{black}Figure}
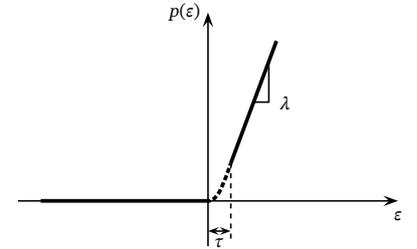
\begin{wrapfigure}{r}{0.33\textwidth}
	\footnotesize
	\vspace*{-0.5cm}
	\begin{center}
		\begin{pspicture}(10,4)(62,35)
		% coordinate system
		\psline[linewidth=0.6pt,arrowsize=4pt]{->}(10,10)(60,10)  % x
		\psline[linewidth=0.6pt,arrowsize=4pt]{->}(35,4)(35,35)  % y
		\rput[tc](60,8){$\varepsilon$}
		\rput[mr](34,35){$p(\varepsilon)$}
		
        % penalty function
		\psline[linewidth=1.4pt](13,10)(35,10)
        \psline[linewidth=1.4pt](38,15)(44,31.2)
        \psbcurve[linewidth=1.4pt,linestyle=dashed,dash=2pt 1pt](35,10)(36,10.5)(37,12.3)(38,15)
        % tolerance
        \psline[linewidth=0.6pt,linestyle=dashed,dash=2pt 2pt]{-}(38,5)(38,15)
        \psline[linewidth=0.6pt,arrowsize=3pt]{<->}(35,6)(38,6)
        \rput[tc](36.5,4.5){$\tau$}
        % slope
        \psline[linewidth=0.6pt]{-}(41,23.1)(43,23.1)(43,28.5)
        \rput[ml](44.5,23){$\lambda$}
 		\end{pspicture}
	\end{center}
	\vspace*{-0.4cm}
	\caption{Penalty function $p(\varepsilon)$ for the violation $\varepsilon$ of the state constraints. Within a tolerance region defined by $\tau>0$, the kink is approximated by a cubic polynomial (dashed bold line).}
	\label{Fig:SoftConstraints}
    \vspace*{-0.5cm}
  \end{wrapfigure}
\textbf{State Constraints.} The state constraints come in the form of two half-spaces parallel to each trajectory segment, as discussed above and illustrated in Figure \ref{Fig:StateConstraints}. They are included in the FTOCP as \emph{soft constraints}, to avoid infeasibility of the optimization program \cite{ScoRaw:1999}. Soft constraints penalize the amount of \emph{constraint violation} $\varepsilon$, by means of adding a \emph{penalty function} $p(\varepsilon)$ to the objective function of the FTOCP, instead of strictly enforcing the constraints at all times.

The constraint violation $\varepsilon$ amounts to the distance with which the vehicle violates the, left or right, state constraint (in meters). A common shape for the penalty function is to be zero in the feasible region ($\varepsilon<0$) and to increase linearly in the infeasible region ($\varepsilon>0$). However, this causes a kink at $\varepsilon=0$, and hence a lack of differentiability of the objective function. Therefore, in \autompc, the kink is interpolated by a cubic polynomial within a small tolerance inside the infeasible region, as shown in Figure \ref{Fig:SoftConstraints}. As a result, the penalty function is twice continuously differentiable.

The slope of the penalty function $\lambda>0$ is adjustable during runtime. The size of the interpolation region is defined by a parameter $\tau>0$. For notational simplicity, the functions $p_{\mathrm{left},k}:\BR^{n}\to\BR$ and $p_{\mathrm{right},k}:\BR^{n}\to\BR$ are used to return the corresponding penalty values for the left and right state constraint violations. Thus, $p_{\mathrm{left},k}\bigl(z_{k}\bigr)$ and $p_{\mathrm{right},k}\bigl(z_{k}\bigr)$ return the penalty incurred by the state of the vehicle $z_{k}$ in prediction step $k$ from the left and right state constraint, respectively.

\begin{inst}[frametitle=How to Define the Cost Function and the Constraints]
  \textbf{Cost function.} The input weights $r_{1},r_{2},\dots,r_{m}$ and the state weights $q_{1},q_{2},\dots,q_{n}$ in \eqref{Equ:InputCostFunction} and \eqref{Equ:StateCostFunction} are external inputs to the \autompc\; code. They can hence be tuned on-line, during the operation of the controller. They are defined via an $m$-dimensional vector, called \texttt{R}, and an $n$-dimensional vector, called \texttt{Q}. The elements of these vectors are the input and state weights in the orders corresponding to $z_{k}$ and $u_{k}$:
  \begin{equation}\label{Equ:InputStateWeights}
    \texttt{R}\triangleq
    \begin{bmatrix}
      r_{1}\\
      r_{2}\\
      \vdots\\
      r_{m}
    \end{bmatrix}\ec\qquad
    \texttt{Q}\triangleq
    \begin{bmatrix}
      q_{1}\\
      q_{2}\\
      \vdots\\
      q_{n}
    \end{bmatrix}\ef
  \end{equation}
  Recall that all input weights must be selected as positive scalars, $r_{1},r_{2},\dots,r_{m}>0$, and all state weights must be selected as non-negative scalars, $q_{1},q_{2},\dots,q_{n}\geq 0$.

  \textbf{Input constraints.} The input bounds and rate constraints are defined by an $4m$-dimensional vector, called \texttt{Ucon}, listing (1) the lower bounds, (2) the upper bounds, (3) the lower rate limits, (4) the upper rate limits:
  \begin{equation}
	U_{\min}\triangleq
    \begin{bmatrix}
      a_{\min}\\
      \gamma_{\min}\\
      u_{\min}(3)\\
      \vdots\\
      u_{\min}(m)
    \end{bmatrix}\ec\qquad
    U_{\max}\triangleq
    \begin{bmatrix}
      a_{\max}\\
      \gamma_{\max}\\
      u_{\max}(3)\\
      \vdots\\
      u_{\max}(m)
    \end{bmatrix}\ec\qquad
    \Delta U_{\min}\triangleq
    \begin{bmatrix}
      \Delta a_{\min}\\
      \Delta \gamma_{\min}\\
      \Delta u_{\min}(3)\\
      \vdots\\
      \Delta u_{\min}(m)
    \end{bmatrix}\ec\qquad
    \Delta U_{\max}\triangleq
    \begin{bmatrix}
      \Delta a_{\max}\\
      \Delta \gamma_{\max}\\
      \Delta u_{\max}(3)\\
      \vdots\\
      \Delta u_{\max}(m)
    \end{bmatrix}\ef
  \end{equation}
  The elements are as defined in \eqref{Equ:InputBounds} and \eqref{Equ:InputRates}. All intervals must be non-empty and contain $0$; thus all lower bounds and rate limits must be negative and all upper bounds and rate limits must be positive. \texttt{Ucon} is then constructed by stacking up the above vectors:
  \begin{equation}
    \texttt{Ucon}\triangleq
    \begin{bmatrix}
      U_{\min}\\
      U_{\max}\\
      \Delta U_{\min}\\
      \Delta U_{\max}
    \end{bmatrix}\ef
  \end{equation}
  The vector \texttt{Ucon} is also an external input to the \autompc\; code. Hence the input constraints can be adjusted on-line, but remain constant over the prediction horizon within a single instance of the FTOCP.

  \textbf{State constraints.} The main part of the state constraints are defined via the maximum deviations $d_{\mathrm{left}}^{(i)}$ and $d_{\mathrm{right}}^{(i)}$ to the left and the right of the reference trajectory. The slope of the penalty function and the tolerance region can be adjusted by two external inputs to the \autompc\; code, called \texttt{conpenalty} ($\lambda$) and \texttt{contolerance} ($\tau$). Both parameters must be positive. The value of \texttt{conpenalty} ($\lambda$) should be such that the state constraints receive priority over the remaining terms in the cost function. The value of \texttt{contolerance} ($\tau$) should be small. A reasonable safety margin should be included in the state constraints, as a compensation for noise and model uncertainty.
\end{inst}

\subsection{Finite-time Optimal Control Problem (FTOCP)}

The overall \emph{Finite-time Optimal Control Problem} (FTOCP) to be solved by the MPC is hence combined as follows:
\begin{subequations}\label{Equ:MPCProblem}\begin{align}
	\text{min}\quad &\sum_{i=1}^{N}\ell_{\mathrm{u}}\bigl(u_{k-1}\bigr)+\ell_{\mathrm{z}}\bigl(z_{k}\bigr)+p_{\mathrm{left}}\bigl(z_{k}\bigr)+
	p_{\mathrm{right}}\bigl(z_{k}\bigr)\\
	\text{s.t.}\quad&z_{k+1}=f\bigl(z_{k},u_{k}\bigr)\ec\hspace*{2.45cm}\fa\;k=0,1,\dots,N-1\ec\\
	&U_{\min}\leq u_{k}\leq U_{\max}\ec\hspace*{2.22cm}\fa\;k=0,1,\dots,N-1\ec\\
    &\Delta U_{\min}\leq \frac{1}{t_{\mathrm{s}}}\bigl(u_{k}-u_{k-1}\bigr)\leq\Delta U_{\max}\ec\quad\fa\;k=0,1,\dots,N-1\ef
\end{align}\end{subequations}
The solution to the FTOCP at time step $k$ provides the \emph{optimal} control inputs over the prediction horizon, which are denoted with a star:
\begin{equation}
  U_{k}\opt\triangleq
  \begin{bmatrix}
      u_{0,k}\opt\\
      u_{1,k}\opt\\
      \vdots\\
      u_{N-1,k}\opt
  \end{bmatrix}\in\BR^{Nm}\ef
  \end{equation}
  The additional time index  $i=0,1,\dots,N-1$ in $u_{i,k}\opt$ refers to the time step $k+i$ for which the optimal control input is intended, i.e., $u_{k+i}=u_{i,k}\opt$, as calculated in time step $k$.

  \textbf{Initial condition.} The MPC model (\ref{Equ:MPCProblem}b) comes with an appropriate initial condition $z_{0}$. It is given by the measurement of the current state of the vehicle, or alternatively by an estimate of that state. Without further measures, the first optimal control input $u_{0,k}\opt$, from the solution to \eqref{Equ:MPCProblem}, is applied to the vehicle. Note that this assumes that the computation time for solving \eqref{Equ:MPCProblem} is very small (compared to $t_{\mathrm{s}}$). Also further latencies in the functional chain (due to measurement data processing, state estimation, communication delays, etc.) must be negligible. This approach can be realized in \autompc by choosing the code parameter \texttt{onesteppred} as `0'.
  
  Since this approach is not always realistic, another option is to set the code parameter \texttt{onesteppred} to `1'. In this case, the MPC does \emph{not} use the measurement (or estimate) of the current state of the vehicle as the initial condition $z_{0}$. Instead, it is replaced by a calculated prediction over one time step, using the internal model of the MPC. This means, essentially, the MPC always operates one step ahead in the future. Hence the control input $u_{0,k}\opt$ from its solution should be applied in the time step after solving \eqref{Equ:MPCProblem}. This approach provides one extra time step to account for the computations and other latencies in the functional chain. 
  
\begin{inst}[frametitle=Outputs of the FTOCP]
  The outputs of \autompc\; code include the first optimal control input $u_{0,k}\opt$, which should be applied to the system. Moreover, the full sequence of optimal control inputs, $U_{k}\opt$, can be gathered in each time step $k$. The corresponding optimal state sequence $Z_{k}\opt$, according to the prediction model (\ref{Equ:MPCProblem}b) in the FTOCP, is provided as an additional output (for debugging purposes):
  \begin{equation}
  Z_{k}\opt\triangleq
  \begin{bmatrix}
      z_{0,k}\opt\\
      z_{1,k}\opt\\
      \vdots\\
      z_{N,k}\opt
  \end{bmatrix}\in\BR^{(N+1)n}\ef
  \end{equation}

  The \autompc\; code also indicates the target \emph{driving mode}, called \texttt{drivmode}, e.g., for the purpose of gear shifts or for operating the parking brake. The driving mode may assume three possible values:
  \vspace*{-0.15cm}\begin{itemize}[leftmargin=.35cm,labelsep=.2cm]
    \item $\texttt{drivmode}=0$: standstill
    \item $\texttt{drivmode}=1$: forward driving
    \item $\texttt{drivmode}=2$: reverse driving
  \end{itemize}\vspace*{-0.15cm}
  Note that changes in direction are not necessarily identical to those demanded in the reference path or trajectory. This is because \autompc\; also accounts for dynamics when generating the target driving mode. In particular, changing between forward and reverse driving is only possible when the vehicle is at rest, thus providing an extra level of safety. For example, the \autompc\; will not allow a shift into the reverse gear while the vehicle is driving forward at high speeds. If a reversing trajectory were provided at times of forward driving, \autompc\; proceeds with braking the vehicle to a standstill before actually reflecting the desired direction change.
  
  As additional information (for debugging purposes), the reference values used in the cost function of the FTOCP are provided as another output of the \autompc\; block. The output, called \texttt{Ref}, is a vector of dimension $9N$, containing the stacked-up reference values for each prediction step $k=1,2,\dots,N$:
  \begin{equation}
  \texttt{Ref}\triangleq
  \begin{bmatrix}
    \texttt{ref}_{1}\\
    \texttt{ref}_{2}\\
    \vdots\\
    \texttt{ref}_{N}
  \end{bmatrix}\;\in\BR^{9N}\ec\qquad
  \texttt{ref}_{k}\triangleq
   \begin{bmatrix}
    x^{\mathrm{(ref)}}_{k}\\
    y^{\mathrm{(ref)}}_{k}\\
    \varphi^{\mathrm{(ref)}}_{k}\\
    v^{\mathrm{(ref)}}_{k}\\
    a^{\mathrm{(ref)}}_{k}\\
    \delta^{\mathrm{(ref)}}_{k}\\
    \beta^{\mathrm{(ref)}}_{k}\\
    d_{\mathrm{left},k}^{\mathrm{(ref)}}\\
    d_{\mathrm{right},k}^{\mathrm{(ref)}}
  \end{bmatrix}\;\in\BR^{9}\ef
  \end{equation}
  \textbf{Summary.} The following is the list of outputs of the \autompc\; code, in the actual order:
  \vspace*{-0.15cm}\begin{itemize}[leftmargin=.55cm,labelsep=.2cm]
     \item[(1)] target driving mode $\texttt{drivmode}\in\{0,1,2\}$;
     \item[(2)] optimal control input $u_{0,k}\opt\in\BR^{m}$;
     \item[(3)] optimal input sequence $U_{k}\opt\in\BR^{Nm}$;
     \item[(4)] reference values for the cost function $\texttt{Ref}\in\BR^{9N}$.
     \item[(5)] optimal state sequence $Z_{k}\opt\in\BR^{(N+1)n}$;
  \end{itemize}\vspace*{-0.15cm}
\end{inst}

\section{Nonlinear Active Set (NAS) Method}

%- one-step prediction

%- Theorem: Persistent feasibility

%- Theorem: Convergence to local minimum

The Nonlinear Active Set (NAS) method used to solve the FTOCP \eqref{Equ:MPCProblem} is essentially the same as previously presented \cite{Schildi:2016}.

\textbf{Initialization.} The iterations of the NAS method are counted with $j=1,2,\dots$ (not to be confused with the numbering of vector elements above). Hence $U^{(j)}$ denotes the new input sequence computed in iteration $j$. For simplicty, the closed-loop time index $k$ is omitted in the notation of this section, assuming that the NAS is described in time step $k$.

A major advantage of the NAS method is the possiblilty of \emph{warm starting}. Thus in step $k$, the solver is initialized with the shifted solution of the previous time step $k-1$, if it exists and it is feasible:
\begin{equation}
  U^{(0)}\leftarrow
  \begin{bmatrix}
      u_{1,k-1}\opt\\
      u_{2,k-1}\opt\\
      \vdots\\
      u_{N-1,k-1}\opt\\
      u_{N-1,k-1}\opt
  \end{bmatrix}\ef
\end{equation}
If a previous solution does not exist, such as in the very first step, the solver is initialized with an input sequence of all zeros.\footnote{Choosing all inputs equal to zero is guaranteed to be feasible, by the definition of the input constraints and the lack of state constraints.} If the previous solution is not feasible, e.g., due to a change in the input constraints, it is projected onto the feasible set.

\textbf{Iterations.} The algorithm starts from $U^{(0)}$ and the corresponding initial set of active constraints $\mathcal{A}^{(0)}$. The input sequence and the corresponding state sequence after each iteration $j$ are denoted
\begin{equation}
  U^{(j)}\triangleq
  \begin{bmatrix}
    u_{0}^{(j)}\\
    u_{1}^{(j)}\\
    \vdots\\
    u_{N-1}^{(j)}
  \end{bmatrix}\in\BR^{Nm}\qquad\text{and}\qquad
  X^{(j)}\triangleq
  \begin{bmatrix}
    x_{0}^{(j)}\\
    x_{1}^{(j)}\\
    \vdots\\
    x_{N}^{(j)}
  \end{bmatrix}\in\BR^{N(n+1)}\ec
\end{equation}
respectively. The index $k$ in $u_{k}^{(j)}$ and $x_{k}^{(j)}$ now refers to the \emph{prediction step} inside the FTOCP. The \emph{set of active constraints} at iteration $j$ is a subset of the constraints (\ref{Equ:MPCProblem}c,d) in the FTOCP,
\begin{equation}
\mathcal{A}^{(j)}\subset\Bigl\{1,2,\dots,(4N-2)m\Bigr\}\ef
\end{equation}
There are $Nm$ lower and upper input bounds, respectively, and $(N-1)m$ lower and upper rate constraints, respectively. Thus the FTOCP contains $(4N-2)m$ constraints in total.\footnote{The first rate constraints, for $k=0$ in (\ref{Equ:MPCProblem}d), is effectively a bound on the input $u_{0}$.} For simplicity, the following more general notation will be used: 
\begin{equation*}
g_{l}\bigl(U^{(j)}\bigr)\leq 0\ec\quad\text{where}\;
g_{l}\bigl(U^{(j)}\bigr)= 0\;\;\text{if and only if}\;\;l\in\mathcal{A}^{(j)}\ef
\end{equation*}
The first step in each iteration $j$ is to find the \emph{search direction} $\tilde{U}^{(j)}\in\BR^{Nm}$, 
\begin{equation}
  \tilde{U}^{(j)}\triangleq
  \begin{bmatrix}
    \tilde{u}_{0}^{(j)}\\
    \tilde{u}_{1}^{(j)}\\
    \vdots\\
    \tilde{u}_{N-1}^{(j)}
  \end{bmatrix}\ef
\end{equation}
The search direction is then used to determine the next iterate by a \emph{step size} $\alpha>0$, based on a \emph{backtracking line search},
\begin{equation}\label{Equ:NewIterate}
U^{(j+1)}=U^{(j)}+\alpha\cdot\tilde{U}^{(j)}\ef
\end{equation}

\textbf{Search Direction.} The search direction in iteration $j$ is computed as the solution to the following \emph{Local Quadratic Program} (Local QP):
  \begin{subequations}\label{Equ:LocalQP}\begin{align*}
	\hspace*{-0.8cm}\text{min}\quad &\sum_{k=0}^{N-1}\tilde{u}_{k}^{(j)\scriptsize{\mathrm{T}}} \tilde{R}\tilde{u}_{k}^{(j)} +
	\tilde{z}_{k+1}\tp \tilde{Q}\tilde{z}_{k+1}+e_{k}\tp\tilde{u}_{k}^{(j)}+f_{k+1}\tp\tilde{z}_{k+1}\\
	\hspace*{-0.8cm}\text{s.t.}\quad&\tilde{z}_{k+1}=A_{k}\tilde{z}_{k}+B_{k}\tilde{u}_{k}^{(j)}\qquad
	\fa\;k=0,\dots,N-1\ec\\
	&g_{l}\bigl(\tilde{U}^{(j)}\bigr)=0\qquad\fa\;l\in\mathcal{A}^{(j)}\ef
  \end{align*}\end{subequations}
Here $A_{k}\in\BR^{n\times n}$ and $B_{k}\in\BR^{n\times m}$ represent the \emph{linearization} of the nonlinear model \eqref{Equ:DiscreteModel} around $U^{(j)}$ and $Z^{(j)}$. The linearization is obtained numerically, by taking \emph{finite differences} within the nonlinear simulation model (code parameter: \texttt{finitediff}). The linear cost vectors $e_{k}\in\BR^{m}$ and $f_{k+1}\in\BR^{n}$ are defined accordingly, by a linearization of the cost function (\ref{Equ:MPCProblem}a) around $U^{(j)}$ and $Z^{(j)}$. The matrices $\tilde{R}\in\BR^{m\times m}$ and $\tilde{Q}\in\BR^{n\times n}$ are diagonal, with the input and state weights in \eqref{Equ:InputStateWeights} as the diagonal elements.

\textbf{Karush-Kuhn-Tucker (KKT) Conditions.} Note that the Local QP \eqref{Equ:LocalQP} is strictly convex and has only equality constraints. Thus it has a unique solution, which is characterized by its corresponding KKT conditions (\cite[Sec.\,5.5]{BoydVan:2004}). Due to the absence of inquality constriants, the KKT conditions constitute a system of linear equations, with a particular sparsity structure. Hence it can be solved by an approach similar to the one proposed in (\cite{Domahidi:2013}). 

To this end, define the vector of stage-wise decision variables $z\in\BR^{N(m+n)}$ as
\begin{equation}
  \xi\triangleq
  \begin{bmatrix}
    \tilde{u}_{0}\\
    \tilde{z}_{1}\\
    \tilde{u}_{1}\\
    \tilde{z}_{2}\\
    \vdots\\
    \tilde{u}_{N-1}\\
    \tilde{z}_{N}
  \end{bmatrix}\ef
\end{equation}
The equality constraints on $\xi$ imposed in the local QP can be expressed as
\begin{equation}
\underbrace{\begin{bmatrix}
	D_{1} & &\dotsm &0 \\
	C_{1} &D_{2} & &\vdots  \\
	\vdots &\ddots &\ddots &\\
	0 &\dotsm &C_{N-1} &D_{N}
	\end{bmatrix}}_{\triangleq G}
\xi=
\underbrace{\begin{bmatrix}
	0\\
	0\\
	\vdots\\
	0
	\end{bmatrix}}_{\triangleq h}\ec
\end{equation}
where
\begin{equation*}
D_{k}\triangleq
\underbrace{\begin{bmatrix}
	E_{k} & O_{a_{k}\times n} \\
	B_{k} & -I_{n\times n}
	\end{bmatrix}}_{\in\BR^{(a_{k}+n)\times (m+n)}}\ec\quad
C_{k}\triangleq
\underbrace{\begin{bmatrix}
	F_{k+1} & O_{a_{k+1}\times n} \\
	O_{n\times m} & A_{k+1}
	\end{bmatrix}}_{\in\BR^{(a_{k+1}+n)\times (m+n)}}\ef
\end{equation*}
Here $O_{a_{k}\times n}\in\BR^{a_{k}\times n}$ denotes the zero matrix and $I_{n\times n}\in\BR^{n\times n}$ denotes the identity matrix. The matrices $E_{k},F_{k}\in\BR^{a_{k}\times m}$ contain only zeros and ones, according to the active bounds and rate constraints. The numbers $a_{k}$ denote the number of active constraints (bounds and rates) pertaining to the inputs in stage $k$. Thus $G\in\BR^{(Nn+a)\times N(m+n)}$ and $h\in\BR^{Nn+a}$, where $a\triangleq\sum_{k=1}^{N}a_{k}$. Taking proper care with the setup of the active set, and avoiding redundancies, all $a_{k}$ become no larger than $m$, and $G$ retains full row rank.

Using the above definitions, the KKT optimality conditions for the local QP can expressed as (\cite[Ch.\,16]{NocWri:2006})
\begin{equation}\label{Equ:KKT}
\begin{bmatrix}
H & G\tp \\
G & O_{a\times a}
\end{bmatrix}
\begin{bmatrix}
\xi \\
\nu 
\end{bmatrix}
=
\begin{bmatrix}
-f \\
0 
\end{bmatrix}\ec
\end{equation}
where $H\in\BR^{N(m+n)\times N(m+n)}$ is the collective Hessian matrix and $f\in\BR^{N(m+n)}$ is the collective linear term of the Local QP. Instead of solving \eqref{Equ:KKT} directly for $\xi$ and the \emph{Lagrangian multipliers} $\nu$, the solution proceeds by block elimination (\cite[Sec.\,5.5]{BoydVan:2004}). As $H$ is positive definite (because all cost weights are positive), substituting the first line of \eqref{Equ:KKT} into the second one gives
\begin{subequations}
\begin{equation}\label{Equ:PosDefSystem}
\xi = -H^{-1}f-H^{-1}G\tp\nu
\qquad\Longrightarrow\qquad GH^{-1}G\tp\nu = -GH^{-1}f\ef
\end{equation}\end{subequations}
In a first step, the equation on the right-hand side of \eqref{Equ:PosDefSystem} is solved for $\nu$; in a second step, this $\nu$ is substituted into the equation on the left-hand side of \eqref{Equ:PosDefSystem} to obtain $\xi$.

\textbf{Cholesky factorization.} The main computational effort is to factorize the matrix $GH^{-1}G\tp$, which is positive definite, because $H$ is positive definite and $G$ has full row rank, and sparse. A generic \emph{Cholesky factorization} (\cite[Sec.\,2.7.1]{Demmel:1996}) yields a lower-triangular, non-singular, sparse factor $L$ such that
\begin{equation}\label{Equ:Cholesky}
GH^{-1}G\tp = LL\tp\ef
\end{equation}
With $L$ given, $\nu$ can be computed very cheaply. The effort for the factorization itself, and for computing $\nu$ and $\xi$, increases only linearly with the prediction horizon $N$. An iterative refinement procedure (\cite[Sec.\,2.5]{Demmel:1996}) can be used to increase the numerical accuracy (code parameter: \texttt{maxiterref}).

\textbf{Backtracking Line Search.} After computing the search direction $\tilde{U}^{(j)}$, the NAS algorithm performs a line search along $\tilde{U}^{(j)}$. The goal is to find the optimal \emph{step size} $\alpha>0$ that minimizes the actual cost function of the (nonlinear) FTOCP (\ref{Equ:MPCProblem}a), called here $J(\alpha)$. The search for the optimal $\alpha$, the code employs a \emph{backtracking line search} algorithm (\cite[Ch.\,3]{NocWri:2006}).

The line search initiates at an upper bound
\begin{equation}
    \alpha^{(0)}\leftarrow\alpha_{\max}\ec
\end{equation} 
whose maximum value is $1$.\footnote{The reason is that $\alpha=1$ directly leads to the (unique) minimum of the Local QP. Since the Local QP is based on a linearization, which is only valid in some neighborhood, any further decrease of the cost beyond $\alpha=1$ (i.e., for $\alpha>1$) is completely a matter of chance, and thus similar to selecting a random search direction. This, however, is against the rationale of the presented NAS method, which is based on computing a second-order, and hence high-quality, search direction.} However, $\alpha_{\max}\in[0,1]$ can actually be smaller if the line search is blocked by a non-active constraint. Thus $\alpha^{(0)}$ is selected as the largest step size, capped at $1$, that maintains feasibility of the iteration \eqref{Equ:NewIterate}.

The line search proceeds iteratively, with an iteration counter $i=0,1,2,\dots$ (not to be confused with the trajectory segments above), updating $\alpha$ geometrically:
\begin{equation}
  \alpha^{(i+1)}\leftarrow c_{\mathrm{bt}}\alpha^{(i)}\ec
\end{equation}
where $c_{\mathrm{bt}}\in(0,1)$ is an adjustable parameter that reduces the step size in each line search (code parameter: \texttt{backtrack}). The line search terminates if the step size $\alpha^{(i)}$ satisfies the \emph{Armijo condition}:
\begin{equation}\label{Equ:ArmijoCondition}
  J\bigl(\alpha^{(i)}\bigr)-J\bigl(0\bigr)\leq c_{\mathrm{dc}}\,\alpha^{(i)}\left.\frac{\partial J(\alpha)}{\partial \alpha}\right|_{\alpha=0}\ec
\end{equation}
where $c_{\mathrm{dc}}\in(0,1)$ is an adjustable parameter that determines the required decrease of the cost function per step size (code parameter: \texttt{decrease}). In all cases, of course, the \emph{best} (and not the \emph{last}) of all examined step sizes is finally selected, i.e., the one that leads to the highest decrease of the objective function.

\textbf{Projected Newton Steps.} If the line search hits a new active constraint, the search direction is projected onto this constraint. In particular,
\vspace*{-0.15cm}\begin{itemize}[leftmargin=.35cm,labelsep=.2cm]
  \item if the search hits an (upper or lower bound) in a particular input, the corresponding entry in the search direction $\tilde{U}^{(j)}$ is set to zero;\
  \item if the line search hits an (upper or lower) rate constraint, the corresponding two elements of the search direction $\tilde{U}^{(j)}$ are both equated to their average value, or to zero in case a bound is active simultaneously.
\end{itemize}\vspace*{-0.15cm}
The projected search direction is used to continue the line search, instead of computing a new search direction immediately (\cite[Sec.\,16.7]{NocWri:2006}). This approach of Projected Newton methods has also been used for other optimization problems in control (\cite{Bertsekas:1982}, \cite{AxeHans:2008}). The advantage stems from the fact that the computation of the search direction is the computationally most expensive step of the NAS algorithm; the drawback is that the quality of the search direction deteriorates with an increasing number of projections. The maximum number of projections for a search direction can be defined by the code parameter \texttt{maxproj}.

Conversely, an active constraint (bound or rate constraint) is released from the active set if the corresponding Langrange multiplier is negative. To account for numeric noise, the multiplier must be negative by at least a small amount $\nu_{\mathrm{tol}}>0$ (code parameter: \texttt{dualtol}).

\textbf{Termination.} The NAS algorithm terminates if a local optimum is reached, i.e., if the KKT conditions of the Local QP \eqref{Equ:KKT} are satisfied and no further constraint can be released from the active set.

For an early termination, the maximum number of iterations can be restricted to some value $j_{\max}\in\BZ_{+}$ (code parameter: \texttt{maxit}). This can also be used to limit the total runtime of the algorithm, and potentially distribute the computational effort for hard problems over multiple time steps. However, no strict real-time bound can be provided, as the computation time also depends on the simulation model, the selected integration method, and other code parameters.

\begin{inst}[frametitle=How to Select the NAS Algorithm Parameters]
  The following is a list of parameters that have to be specified for the NAS algorithm, with some instructions on how they should be chosen.
  
  \texttt{finitediff}: For calculating derivatives, i.e., rates of change, each entry of $U^{(j)}$ and $Z^{(j)}$ is modified by a \emph{finite difference} to its true value. This parameter can be set via \texttt{finitediff}. It may seem that the smaller the parameter, the more accurate the derivative. However, even in principle, there is no gain by picking its value too small, and it may lead to numerical problems. Appropriate values are in the order of $10^{-4}$ to $10^{-6}$.

   \texttt{maxit}: Fixing the maximum number of NAS iterations limits the (maximum) computation time, but may lead to suboptimal solutions. A key advantage of the NAS algorithm is that primal feasibility is preserved during all iterations, while it monotonically improves on the objective function. Few NAS iterations usually suffice for convergence, especially when a good warm start is available. Moreover, suboptimality is known to have little impact on the closed-loop performance of MPC performance \cite{WangBoyd:2010}. Generally, driving curvier paths at higher speeds, and especially the presence of obstacles, increases the required number of NAS iterations. In general, about 2 to 10 should usually suffice.

   \texttt{maxproj}: The maximum number of projections of the search direction onto a new active set may has a major impact on the computational performance of the NAS algorithm. Suitable values can be expected to increase with the complexity of the driving task, similar to \texttt{maxit}, and the MPC horizon length. To some extent, increasing \texttt{maxit} and \texttt{maxproj} will have a similar effect. Depending on \texttt{maxit},  a proper choice for \texttt{maxproj} may be in the range from 5 to 50.

   \texttt{dualtol}: To release a constraint from the active set, its corresponding multiplier should be negative, in particular less than $-\texttt{dualtol}$. The effect of this parameter on the outcome of the NAS is minor. Hence it should be chosen small, yet somewhat above the machine epsilon, e.g., $10^{-8}$ to $10^{-12}$.

   \texttt{maxiterref}: The number of iterations for the iterative refinement usually plays a minor role for the MPC results. It is not advisable to pick a high number here, i.e., either 0 or 1.
   
   \texttt{backtrack}: Higher values of $c_{\mathrm{bt}}$ (e.g., 0.8) lead to better results for the step size, but come with higher computational cost. Lower values of $c_{\mathrm{bt}}$ (e.g., 0.8) tend to converging to a final step size faster, but possibly missing good step sizes \cite[Ch.~3]{NocWri:2006}. An appropriate range is $[0.1,0.8]$, with $0.5$ being the most commonly used value.
   
   \texttt{decrease}: Lower values of $c_{\mathrm{dc}}$ lead to an easier acceptance of the step size, and thus a quicker line search. Higher values of $c_{\mathrm{dc}}$ demand a higher decrease of the cost function per step size, and thus lead to more line search iterations. While the parameter needs to be selected in combination with $c_{\mathrm{bt}}$, lower values in the range of $10^{-1}$ to $10^{-4}$ have proven to work well in practice.
\end{inst}

\section{ROS Package}

The \acknlmpc\; package constitutes a ROS2 wrapper for \autompc. It is configured for the control of vehicles with Ackermann steering geometry. However, it can be adapted for other vehicles as well. Besides the code for Nonlinear MPC, it provides a simple simulation and several examples for vehicle reference tracking. The package source code and the MPC code generator are available under the following links:\\
\hspace*{2.0cm}\href{https://git.ime.uni-luebeck.de/public-projects/asl/ackermann_nlmpc}{\texttt{https://git.ime.uni-luebeck.de/public-projects/asl/ackermann\_nlmpc}}\\
\hspace*{2.0cm}\href{https://git.ime.uni-luebeck.de/public-projects/asl/autompc}{\texttt{https://git.ime.uni-luebeck.de/public-projects/asl/autompc}}\\
The detailed installation and setup instructions are provided inside the \texttt{readme} and \texttt{launch} files in the respective repositories.

The \texttt{initmpc\_ros.m} script in the \autompc\; package can be used to generate C code compatible with the ROS package. This generated C code is loaded as a shared library at runtime in the \acknlmpc\; package. A version generated with default parameters for a dynamic bicycle model is provided with the package. It can be replaced if parameter or model changes are desired. Changes to the sampling time \textit{dt}, prediction horizon length \textit{Npar} and maximum trajectory length \textit{Nn} need to be reflected in the \texttt{launch} file of the ROS package.

The main node of the package, \texttt{ackermann\_nlmpc\_node}, subscribes to vehicle odometry and steering angle data and publishes vehicle acceleration and steering angle rate commands. The \emph{reference path} or a \emph{reference trajectory} can be provided via a custom \texttt{Trajectory} message that essentially mirrors the description in Chapter \ref{Chap:Reference}. To improve ease of use and compatibility with path planning algorithms, the pose information in the trajectory header and nodes may be provided in any coordinate frame, as long as a transformation to the \texttt{odometry} frame is available via \texttt{tf2}.

To ensure compatibility with standard ROS2 path planners, which output trajectories in the \texttt{Path} message type, a converter node to the custom \texttt{Trajectory} message type is included with the \acknlmpc\; package. Note that a \texttt{Path} message contains only pose and time information for each node. Hence, additional information must be added during the automatic conversion to a \texttt{Trajectory} message. The most intricate one is the \texttt{driving mode} parameter, for which the converter needs to detect direction changes that are implicitly contained inside the \texttt{Path}. This is implemented by means of a simple geometric comparison of successive nodes; then the \texttt{driving mode} parameter of each output node is set accordingly. Configurable default values are used for the remaining fields (reference velocity, acceleration, and path width constraints) of the trajectory nodes.

\section{Examples}

The examples are started by running the script \texttt{initmpc.m}, in which also all relevant parameters can be specified. The particular example is selected by choosing the variable \texttt{selectedpath} according to the desired scenario.

\subsection{Circular Path with 4 Obstacles}

\renewcommand{\figurename}{\color{black}Figure}
\begin{wrapfigure}{r}{0.5\textwidth}
\centering
\begin{pspicture}(-39,-32)(39,32)
    % reference path
    \psline[linewidth=1.4pt,arrowsize=7pt]{*-*}( 29.9,  2.6)( 29.0,  7.8)
    \psline[linewidth=1.4pt,arrowsize=7pt]{-*}( 29.0,  7.8)( 24.6, 17.2)
    \psline[linewidth=1.4pt,arrowsize=7pt]{-*}( 24.6, 17.2)( 17.2, 24.6)
    \psline[linewidth=1.4pt,arrowsize=7pt]{-*}( 17.2, 24.6)(  7.8, 29.0)
    \psline[linewidth=1.4pt,arrowsize=7pt]{-*}(  7.8, 29.0)(  2.6, 29.9)
    \psline[linewidth=1.4pt,arrowsize=7pt]{-*}(  2.6, 29.9)( -2.6, 29.9)
    \psline[linewidth=1.4pt,arrowsize=7pt]{-*}( -2.6, 29.9)( -7.8, 29.0)
    \psline[linewidth=1.4pt,arrowsize=7pt]{-*}( -7.8, 29.0)(-17.2, 24.6)
    \psline[linewidth=1.4pt,arrowsize=7pt]{-*}(-17.2, 24.6)(-24.6, 17.2)
    \psline[linewidth=1.4pt,arrowsize=7pt]{-*}(-24.6, 17.2)(-29.0,  7.8)
    \psline[linewidth=1.4pt,arrowsize=7pt]{-*}(-29.0,  7.8)(-29.9,  2.6)
    \psline[linewidth=1.4pt,arrowsize=7pt]{-*}(-29.9,  2.6)(-29.9, -2.6)
    \psline[linewidth=1.4pt,arrowsize=7pt]{-*}(-29.9, -2.6)(-29.0, -7.8)
    \psline[linewidth=1.4pt,arrowsize=7pt]{-*}(-29.0, -7.8)(-24.6,-17.2)
    \psline[linewidth=1.4pt,arrowsize=7pt]{-*}(-24.6,-17.2)(-17.2,-24.6)
    \psline[linewidth=1.4pt,arrowsize=7pt]{-*}(-17.2,-24.6)( -7.8,-29.0)
    \psline[linewidth=1.4pt,arrowsize=7pt]{-*}( -7.8,-29.0)( -2.6,-29.9)
    \psline[linewidth=1.4pt,arrowsize=7pt]{-*}( -2.6,-29.9)(  2.6,-29.9)
    \psline[linewidth=1.4pt,arrowsize=7pt]{-*}(  2.6,-29.9)(  7.8,-29.0)
    \psline[linewidth=1.4pt,arrowsize=7pt]{-*}(  7.8,-29.0)( 17.2,-24.6)
    \psline[linewidth=1.4pt,arrowsize=7pt]{-*}( 17.2,-24.6)( 24.6,-17.2)
    \psline[linewidth=1.4pt,arrowsize=7pt]{-*}( 24.6,-17.2)( 29.0, -7.8)
    \psline[linewidth=1.4pt,arrowsize=7pt]{-*}( 29.0, -7.8)( 29.9, -2.6)
    \psline[linewidth=1.4pt,arrowsize=7pt]{-*}( 29.9, -2.6)( 29.9,  2.6)

    % inner constraints
    \psline[linewidth=1.0pt,linestyle=dashed,dash=2pt 2pt]{-}( 32.9,  2.9)( 26.1,  7.0)( 22.1, 15.5)( 15.5, 22.1)(  7.0, 26.1)(  1.8, 20.9)( -1.8, 20.9)( -7.0, 26.1)(-15.5, 22.1)(-22.1, 15.5)(-26.1,  7.0)(-32.9,  2.9)(-32.9, -2.9)(-26.1, -7.0)(-22.1,-15.5)(-15.5,-22.1)( -7.0,-26.1)( -1.8,-20.9)(  1.8,-20.9)(  7.0,-26.1)( 15.5,-22.1)( 22.1,-15.5)( 26.1, -7.0)( 32.9, -2.9)( 32.9,  2.9)

    % outer constraints
    \psline[linewidth=1.0pt,linestyle=dashed,dash=2pt 2pt]{-}( 38.9,  3.4)( 31.9,  8.5)( 27.0, 18.9)( 18.9, 27.0)(  8.5, 31.9)(  2.4, 26.9)( -2.4, 26.9)( -8.5, 31.9)(-18.9, 27.0)(-27.0, 18.9)(-31.9,  8.5)(-38.9,  3.4)(-38.9, -3.4)(-31.9, -8.5)(-27.0,-18.9)(-18.9,-27.0)( -8.5,-31.9)( -2.4,-26.9)(  2.4,-26.9)(  8.5,-31.9)(18.9,-27.0)( 27.0,-18.9)( 31.9, -8.5)( 38.9, -3.4)( 38.9,  3.4)

    % obstacles
    \pspolygon[fillcolor=taucolorinst,linestyle=solid,linewidth=1.0pt,linecolor=darkgray,fillstyle=hlines,hatchwidth=0.2pt,hatchsep=5pt,hatchangle=40](27,2.8)(27,-2.8)(33,-2.8)(33,2.8)
    \pspolygon[fillcolor=taucolorinst,linestyle=solid,linewidth=1.0pt,linecolor=darkgray,fillstyle=hlines,hatchwidth=0.2pt,hatchsep=5pt,hatchangle=40](-27,2.8)(-27,-2.8)(-33,-2.8)(-33,2.8)
    \pspolygon[fillcolor=taucolorinst,linestyle=solid,linewidth=1.0pt,linecolor=darkgray,fillstyle=hlines,hatchwidth=0.2pt,hatchsep=5pt,hatchangle=40](2.8,27)(-2.8,27)(-2.8,33)(2.8,33)
    \pspolygon[fillcolor=taucolorinst,linestyle=solid,linewidth=1.0pt,linecolor=darkgray,fillstyle=hlines,hatchwidth=0.2pt,hatchsep=5pt,hatchangle=40](2.8,-27)(-2.8,-27)(-2.8,-33)(2.8,-33)

    % angles
    \pscircle[fillcolor=black,fillstyle=solid,linewidth=0.1pt,linecolor=black](0,0){0.5}
    \rput(-2,-2){C}
    \psline[linewidth=0.8pt,linestyle=dashed,dash=1pt 1pt]{-}(0,0)( 29.0,  7.8)
    \psline[linewidth=0.8pt,,linestyle=dashed,dash=1pt 1pt]{-}(0,0)( 29.9,  2.6)
    \psline[linewidth=0.8pt,,linestyle=dashed,dash=1pt 1pt]{-}(0,0)( 29.9, -2.6)
    \psline[linewidth=0.8pt,linestyle=dashed,dash=1pt 1pt]{-}(0,0)( 29.0,  -7.8)
    \rput(23,0){$\alpha$}
    \rput(23,4){$\beta$}
    \rput(23,-4){$\beta$}

\end{pspicture}
\caption{Schematic representation of the Circular Path. The reference path (solid line) is centered at C and passes through four obstacles (hatched rectangles). The driving corridor (dashed lines) represents the state constraints of the MPC.}
\label{Fig:ExampleCircular}
\end{wrapfigure}
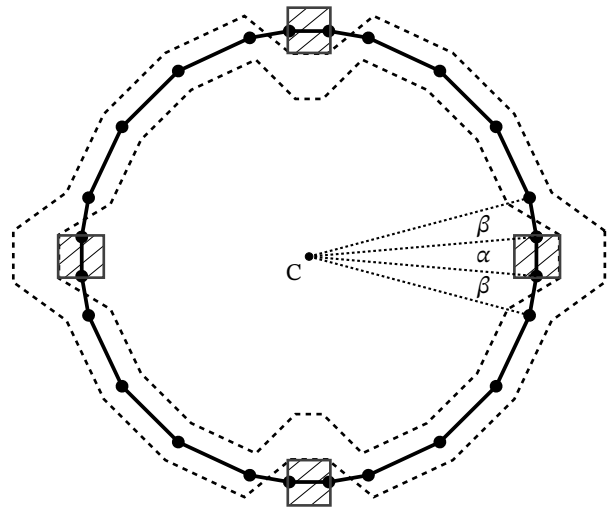

The \emph{Circular Path} example demonstrates the steady cornering of the vehicle and the avoidance of obstacles which intersect with the reference path. Two of the obstacles are avoided to the inside, and two to the outside of the circle; see Figure \ref{Fig:ExampleCircular} for an illustration.

The circle has a variable radius (parameter \texttt{Rad}). It may serve as a test scenario for low and high speed driving, as the reference velocity can be specified by a parameter (\texttt{vref}). It is also possible to choose the driving direction between forward and reverse driving, by using the appropriate path generation scenario (\texttt{selectedpath}).

The size of the obstacles can be specified via the angular size of the obstacles $\alpha$ (parameter \texttt{alpha}) and the angular size of the transition regions $\beta$ (parameter \texttt{beta}). The values are the same for all four obstacles. Besides the obstacle `length', also their `width' can be specified. The two parameters \texttt{Oin} and \texttt{Oout} define the protrusion into the reference path, which deflect the driving corridor. They relate to the two obstacles to be passed on the inside and outside, respectively. The width of the driving corridor can also be adjusted, by the parameter \texttt{pwidth}.

The trajectory type is fixed to `circular path' ($P_{\mathrm{type}}=2$). This means the MPC automatically restarts the circle after reaching its end. The resolution of the reference path, i.e., the number of segments, can be specified by the parameter \texttt{Nn}.

\newpage
\subsection{Forward and Reverse Parking}

\begin{wrapfigure}{r}{0.5\textwidth}
\centering
\vspace*{-0.4cm}
\begin{pspicture}(40,90)
    % reference path
    \psline[linewidth=1.4pt,arrowsize=7pt]{*-*}(5,0)(5,20)
    \psline[linewidth=1.4pt,arrowsize=7pt]{-*}(5,20)(6.5,29.3)
    \psline[linewidth=1.4pt,arrowsize=7pt]{-*}(6.5,29.3)(10.7,37.6)
    \psline[linewidth=1.4pt,arrowsize=7pt]{-*}(10.7,37.6)(17.4,44.3)
    \psline[linewidth=1.4pt,arrowsize=7pt]{-*}(17.4,44.3)(25.7,48.5)
    \psline[linewidth=1.4pt,arrowsize=7pt]{-*}(25.7,48.5)(35,50)
    \psline[linewidth=1.4pt,arrowsize=7pt]{-*}(35,50)(37.5,50)
    \psline[linewidth=1.4pt,arrowsize=7pt]{-*}(35,50)(27.3,51.2)
    \psline[linewidth=1.4pt,arrowsize=7pt]{-*}(27.3,51.2)(20.3,54.8)
    \psline[linewidth=1.4pt,arrowsize=7pt]{-*}(20.3,54.8)(14.8,60.3)
    \psline[linewidth=1.4pt,arrowsize=7pt]{-*}(14.8,60.3)(11.2,67.3)
    \psline[linewidth=1.4pt,arrowsize=7pt]{-*}(11.2,67.3)(10,75)
    \psline[linewidth=1.4pt,arrowsize=7pt]{-*}(10,75)(10,85)
    
    % right constraints
    \psline[linewidth=1.0pt,linestyle=dashed,dash=2pt 2pt]{-}(10,0)(10,20)(11.2,27.7)(14.8,34.7)(20.3,40.2)(27.3,43.8)(35,45)(37.5,45)(25.7,46.5)(17.4,50.7)(10.7,57.4)(6.5,65.7)(5,75)(5,85)
    
    % left constraints
    \psline[linewidth=1.0pt,linestyle=dashed,dash=2pt 2pt]{-}(0,0)(0,20)(1.7,30.8)(6.7,40.6)(14.4,48.3)(24.2,53.3)(35,55)(37.5,55)(28.8,56)(23.2,58.8)(18.8,63.2)(16,68.8)(15,75)(15,85)

    % numbers
    \rput(3,-2){\color{MedGreen}A}
    \rput(40,53){\color{MedBlue}B}
    \rput(7.5,87){\color{MedRed}C}

    % arrows
    \psline[linecolor=MedGreen,linewidth=1.5pt,arrowsize=5pt]{->}(5,0)(5,8)
    \psline[linecolor=MedBlue,linewidth=1.5pt,arrowsize=5pt]{->}(37.5,50)(45.5,50)
    \psline[linecolor=MedRed,linewidth=1.5pt,arrowsize=5pt]{->}(10,85)(10,77)

    % dimensions
    \psline[linewidth=0.8pt,linestyle=dashed,dash=1pt 1pt]{-}(5,0)(37.5,0)
    \psline[linewidth=0.8pt,linestyle=dashed,dash=1pt 1pt]{-}(5,20)(37.5,20)
    \psline[linewidth=0.8pt]{<->}(37.5,0)(37.5,19.5)
    \rput(40,10){$l_1$}

    \psline[linewidth=0.8pt,linestyle=dashed,dash=1pt 1pt]{-}(37.5,20)(37.5,50)
    \pscircle[fillcolor=black,fillstyle=solid,linewidth=0.1pt,linecolor=black](37.5,20){0.5}
    \rput(41,19){$\text{C}_{1}$}
    \psline[linewidth=0.8pt]{<->}(37.2,20.3)(17.7,44)
    \rput(31,32){$r_1$}

    \pscircle[fillcolor=black,fillstyle=solid,linewidth=0.1pt,linecolor=black](37.5,75){0.5}
    \rput(41,74){$\text{C}_{2}$}
    \psline[linewidth=0.8pt,linestyle=dashed,dash=1pt 1pt]{-}(37.5,50)(37.5,75)
    \psline[linewidth=0.8pt]{<->}(15.1,60.6)(37.2,74.7)
    \rput(30,67){$r_2$}

    \psline[linewidth=0.8pt,linestyle=dashed,dash=1pt 1pt]{-}(10,75)(37.5,75)
    \psline[linewidth=0.8pt,linestyle=dashed,dash=1pt 1pt]{-}(10,85)(37.5,85)
    \psline[linewidth=0.8pt]{<->}(37.5,75.5)(37.5,85)
    \rput(40,80){$l_2$}

\end{pspicture}
\caption{Schematic illustration of the Parking example. The reference path (solid line) leads the vehicle from configuration A (green arrow) over configuration B (blue arrow), with a change of driving direction from forward to reverse, to the final configuration C (red arrow). The driving corridor (dashed lines) represents the state constraints of the MPC.}
\label{Fig:ExampleParking}
\end{wrapfigure}
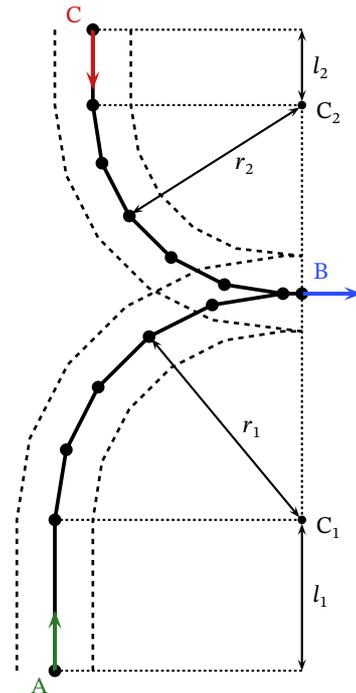

The \emph{Parking} example demonstrates low speed driving as well as direction changes within a reference trajectory. It is available as `forward parking' (starting with a reversing maneuver, then switching to forward parking) and as `reverse parking' (starting with a forward maneuver, then switching to reverse parking). In both cases, the trajectory type is fixed to `path' ($P_{\mathrm{type}}=2$).

Figure \ref{Fig:ExampleParking} shows the case of `reverse parking'. The vehicle first accelerates along a straight segment of length $l_1$ (\texttt{L1}). Then it enters a circular curve, centered at $\text{C}_{1}$ and with radius $r_1$ (\texttt{R1}). At the end of this curve, the vehicle decelerates and finally stops, before performing a direction change, in this case from `forward' to `reverse'. Then it enters into a reverse parking maneuver, consisting first of a circular curve, centered at $\text{C}_{2}$ with radius $r_2$ (\texttt{R2}), and second of a straight segment of length $l_2$ (\texttt{L2}). In point $B$ in Figure \ref{Fig:ExampleParking}, the reference trajectory contains a single segment $i$ with a reference driving mode $D^{(i)}=0$ (standstill). While this is not strictly necessary, it is recommended to include such a standstill node in between direction changes to provide extra time, e.g., for a gear shift or a static steering maneuver.

The maximum velocity can be specified by the scenario parameter (\texttt{vref}). The reference velocity, however, is not constant for this maneuver. The vehicle starts from standstill (in point A) and terminates in standstill (in point C), and also reaches an intermediary standstill during the direction change (in point B). A corresponding reference velocity profile is automatically constructed. It accelerates to \texttt{vref}, then keeps a constant velocity, and finally decelerates back to $0$, in fact twice, for each part of the maneuver. The values for the acceleration and deceleration are taken from the corresponding input bounds. 

The width of the admissible driving corridor can be adjusted by the parameter \texttt{pwidth}. The resolution of the reference path, i.e., the number of segments, can be specified by the parameter \texttt{Nn}.

%----------------------------------------------------------

%\clearpage
\printbibliography

%----------------------------------------------------------

\end{document}